\documentclass[10pt,twocolumn,amssymb,amsmath,nobibnotes,nofootinbib,aps,prb,showpacs,floatfix]{revtex4-2} 
\usepackage{graphicx}
\usepackage{multirow}
\usepackage{eufrak}
\usepackage{bm}
\usepackage{dcolumn}
\usepackage[dvipsnames]{xcolor}
\usepackage{graphicx}
\usepackage{dcolumn}
\usepackage{bm}
\usepackage{soul}
\usepackage{float}
\usepackage[hidelinks=true]{hyperref}
\usepackage[hyphenbreaks]{breakurl}
\hypersetup{
  colorlinks   = true, 
  urlcolor     = blue, 
  linkcolor    = blue, 
  citecolor    = blue 
}
\begin{document}

\title{Semi-Dirac States and Quantum Linear Magnetoresistance in Helimagnetic Pnictide MnP}

\author{Prasanta Chowdhury$^1$, Jyotirmoy Sau$^{2,3}$, Sanat Kumar Adhikari$^{4, 8}$, Sourav Chowdhury$^5$, Peter Bencok$^6$, Matthias Gutmann$^7$, Souvik Chatterjee$^4$, Saurav Giri$^1$, Manoranjan Kumar$^2$, Subham Majumdar$^{1}$    }
\email{sspsm2@iacs.res.in}

\affiliation{$^1$School of Physical Sciences, Indian Association for the Cultivation of Science, 2A \& B Raja S. C. Mullick Road, Jadavpur, Kolkata 700 032, India}

\affiliation{$^2$Department of Condensed Matter and Materials Physics, S. N. Bose National Centre for Basic Sciences, JD Block, Sector III, Salt Lake, Kolkata 700106, India}

\affiliation{$^3$ Department of Physics and Astronomy, Uppsala University, Box-516, S-75120 Uppsala, Sweden}

\affiliation{$^4$ UGC-DAE Consortium for Scientific Research, Kolkata Centre, III, B/4, New Town, Kolkata 700 156, India}

\affiliation{$^5$ Deutsches Elektronen-Synchrotron DESY, Notkestra$\beta$e 85, 22607 Hamburg, Germany}

\affiliation{$^6$ Diamond Light Source, Didcot OX11 0DE, United Kingdom}

\affiliation{$^7$ISIS Neutron and Muon Source, Science and Technology Facilities Council, Rutherford Appleton Laboratory, Chilton Didcot OX11 0QX, United Kingdom}

\affiliation{$^8$Department of Physics, Indian Institute of Technology Kharagpur, West Bengal 721 302, India}

\begin{abstract}
Large linear positive magnetoresistance (LPMR) in topological and magnetic materials remains a subject of intense debate, particularly in noncollinear spin systems where spin-dependent scattering complicates charge transport. Manganese phosphide (MnP), a helimagnetic binary pnictide with multiple field-induced magnetic transitions, provides a useful platform to investigate the interplay between complex magnetism and electronic topology. Here, we present a comprehensive experimental and theoretical investigation of phase-dependent magnetotransport in high-quality MnP single crystals. Hall measurements reveal an anomalous Hall effect dominated by skew scattering at high temperatures and a finite topological Hall effect in the noncollinear fan (FAN) and low-temperature screw (SCR) phases. At low temperatures, we observe a large, non-saturating LPMR reaching nearly 800\% at 4 K and 15 T, with a pronounced linear field dependence in the field-polarized ferromagnetic (FM2) state. First-principles calculations reveal a strongly anisotropic semi-Dirac-like band at the $Y$ point that progressively approaches the Fermi level from the SCR to FAN and FM2 states. Our analysis indicates that the resulting small Fermi pocket can access the extreme quantum-limit regime at experimentally accessible fields, providing a microscopic framework for the observed LPMR within Abrikosov’s quantum magnetoresistance theory.

\end{abstract}

\maketitle

\section{Introduction}

Magnetoresistance (MR) has played a pivotal role in uncovering the topological properties of a wide range of materials. In particular, giant MR has attracted considerable attention owing to its fundamental importance and potential technological applications, including spin-valve sensors~\cite{MR_appl1} and hard disk read/write heads~\cite{MR_appl2}. Numerous topological and semimetallic systems, including WTe$_2$~\cite{WTe2}, NbP~\cite{NbP}, PtBi$_2$~\cite{PtBi2}, Cd$_3$As$_2$~\cite{Cd3As2}, MnBi~\cite{MnBi}, CrP~\cite{CrP_MR}, ZrSiS~\cite{ZrSiS}, CoS$_2$~\cite{CoS2_PNAS}, and many others, exhibit exceptionally large positive MR with either quadratic or nearly linear magnetic-field dependence. While linear positive MR (LPMR) has been widely reported in a variety of topological materials~\cite{CrAs, FeP, MnBi, Cd3As2, Ag2Se,Cd3As2_PRL, TlBiSSe_LMR}, its microscopic origin remains a subject of intense debate. 
\par
In magnetic compounds, large LPMR is relatively uncommon and even rarer in systems with noncollinear spin structures, where additional magnetic scattering (often referred as $s-d$ scattering in transition metal compounds, arising from the interaction of delocalized $s$ electrons with partially localized $d$ electrons) channels tend to reduce carrier mobility and relaxation time. An applied magnetic field tends to suppress the $s-d$ scattering, which is expected to provide a negative MR.   Consequently, understanding the origin of large LPMR in such systems remains an important challenge. MnP, a member of the binary pnictide family (MnP, FeP, CrAs, FeAs, and CrP), provides an ideal platform for this investigation owing to its helical magnetic ground state and multiple field-induced metamagnetic transitions at low temperatures~\cite{MnP_Phase1, MnP_Phase2}. Additionally, noncollinear spin systems can possess a finite scalar spin chirality~\cite{Chirality1,Chirality2}, defined as $\chi_{ijk}=\mathbf{S}_i\cdot(\mathbf{S}_j\times\mathbf{S}_k)$, where $\mathbf{S}_i$, $\mathbf{S}_j$, and $\mathbf{S}_k$ are three neighboring spins, which generates a real-space Berry curvature (BC) that gives rise to the topological Hall effect (THE), as observed in compounds such as SmMn$_2$Ge$_2$, Mn$_5$Si$_3$~\cite{SmMn2Ge2_THE,Mn5Si3_THE}.

\par
MnP-type binary transition-metal pnictides, which crystallize in the orthorhombic \emph{Pnma} structure, exhibit a rich variety of correlated electronic and magnetic phenomena, including pressure-induced superconductivity~\cite{CrAs_SC,MnP_SC}, quantum criticality~\cite{CrAs_QC}, and unconventional magnetism~\cite{FeAs_NM1,FeAs_NM2}. Several members of this family also exhibit large nonsaturating MR arising from distinct microscopic mechanisms. The LPMR observed in CrAs has been attributed to a symmetry-protected semi-Dirac band crossing (characterized by linear dispersion along one momentum direction and quadratic dispersion along the orthogonal direction) in conjunction with a tiny Fermi pocket~\cite{CrAs}, whereas the nonsaturating MR in CrP originates from nearly compensated electron and hole carriers~\cite{CrP_MR}. In contrast, the LPMR reported in FeP is likely due to intra-band scattering in topological bands~\cite{FeP}. The diversity of these mechanisms suggests that the microscopic origin of LPMR is strongly material dependent, thereby motivating a detailed investigation of the magneto-transport properties MnP.
 
\par
\begin{figure}
\centering
\includegraphics[width = 6.5 cm]{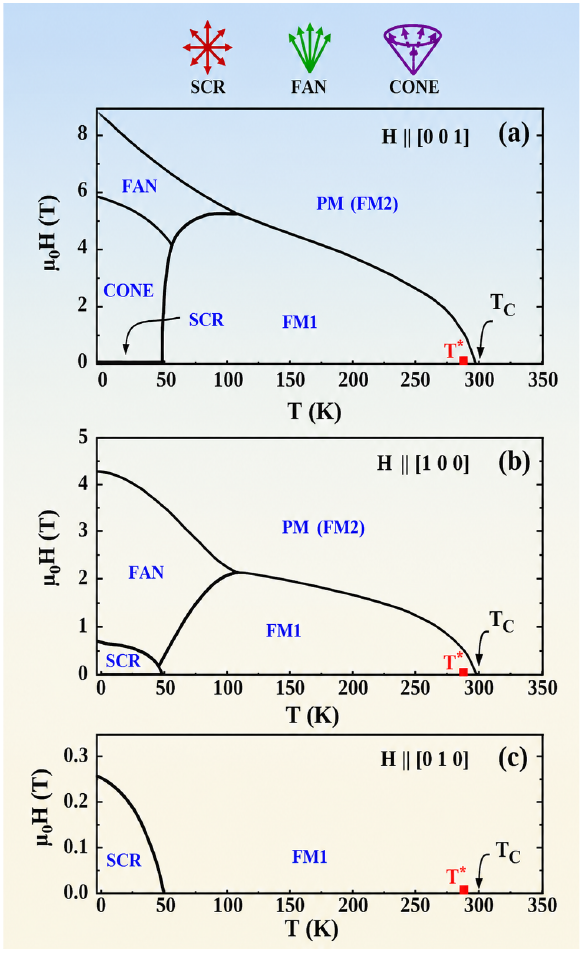}
\caption{\textbf{Phase diagram.} (a), (b) and (c) Shows the standard phase diagram ($\mu_0H$ vs $T$) of MnP for field applied along [0 0 1], [1 0 0] and [0 1 0] direction with different magnetic phase marked.}
\label{fig:Phase}
\end{figure}

MnP is one of the first magnetic materials in which a Lifshitz point~\cite{Lifshitz_Point_MnP_PRL}, where the paramagnetic, ferromagnetic, and helical phases coexist, was experimentally identified, making it a prototypical system for investigating noncollinear magnetism and multicritical magnetic behavior. It exhibits a rich magnetic phase diagram comprising helical, conical, and fan-type spin structures with multiple field- and temperature-induced phase transitions, as summarized schematically in Fig.~\ref{fig:Phase}(a)–(c) for magnetic fields applied along the $c$-, $a$-, and $b$-axes~\cite{MnP_Phase1, Komatsubara,MnP_Phase3,MnP_Phase2}. According to previous neutron diffraction studies~\cite{MnP_Helical,MnP_1980}, in the helimagnetic or screw (SCR) phase, Mn spins rotate within the $ab$-plane with a magnetic propagation vector $\mathbf{q} = (0,~0,~0.112)$, corresponding to a periodicity of roughly nine lattice spacings along the $c$-axis. A second helicoidal phase, known as the fan (FAN) phase, emerges for $H \parallel a$. In this phase, the moments remain confined to the $ab$-plane and oscillate about the $a$-axis, rather than undergoing a full helical rotation, with a propagation vector along the $c$-direction~\cite{MnP_Phase2}. At higher fields for $H \parallel a$, all spins align with the field, resulting in the field-induced ferromagnetic (FM2) phase. The FAN phase is absent for $H \parallel b$. In contrast, for $H \parallel c$, a cone (CONE) phase emerges above the SCR phase, characterized by a finite spin component along the $c$-axis.
\par
Despite the well-established magnetic phase diagram of MnP, how its electronic structure evolves across the SCR, FAN, and FM2 phases remains largely unexplored. In particular, it is unclear how the field-induced reorganization of the Mn moments merely modifies magnetic scattering or also reconstructs the low-energy electronic states governing charge transport. A key question is whether the magnetic configuration can shift band crossings near the Fermi energy ($E_F$) and thereby alter the size and character of the associated Fermi pockets. Establishing this evolution across the distinct magnetic phases is therefore essential for understanding the pronounced phase dependence of the MR, particularly the origin of the large LPMR in MnP as observed in the present study.

\par

In this work, we investigate the magnetic and magnetotransport properties of high-quality MnP single crystals, complemented by first-principles calculations. The low-$T$ MR is strongly phase dependent, culminating in a large, non-saturating LPMR in the high-field FM2 state. Our calculations reveal a semi-Dirac-like band crossing near the $Y$ point, with strongly anisotropic linear and quadratic dispersions that progressively approach the $E_F$ from the SCR through the FAN to the FM2 state. Our analysis indicates that the resulting small Fermi pockets can access the extreme quantum-limit regime at experimentally relevant fields, providing a natural microscopic framework for understanding the observed LPMR within Abrikosov’s theory~\cite{Abrikosov_2000,Abrikosov_1998}. Hall measurements between 100 K and 330 K reveal an AHE dominated by skew scattering, with additional side-jump and intrinsic contributions. Below 100 K, a finite THE is observed in both the FAN and low-$T$ SCR phases, consistent with the previous report in the FAN phase~\cite{MnP_THE}. These results demonstrate a close interplay among the different magnetic states, low-energy electronic structure, and anomalous magnetotransport in MnP.


\begin{figure*}
\centering
\includegraphics[width = 14 cm]{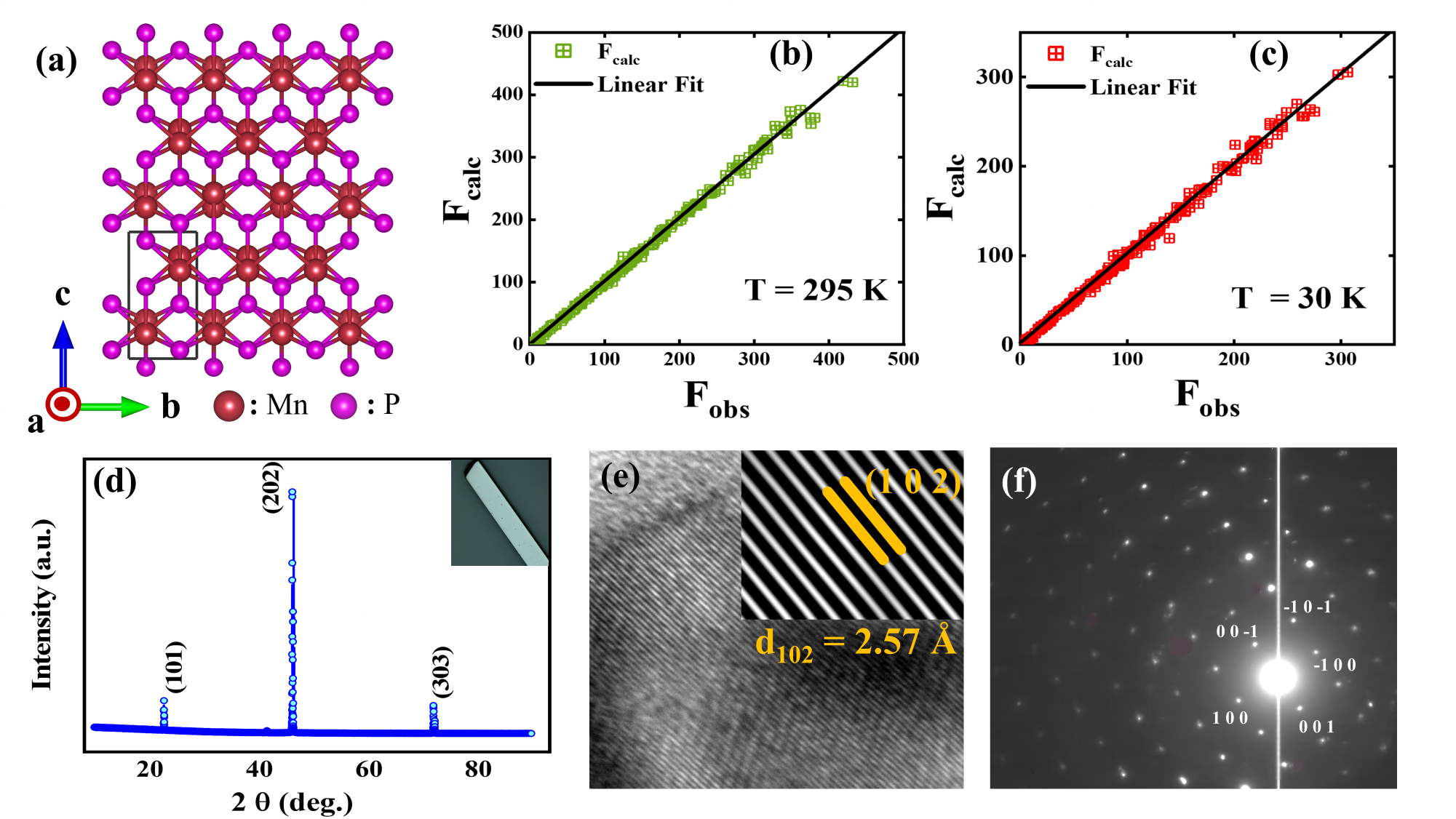}
\caption{\textbf{Structural characterization of single-crystalline MnP.} (a) Crystal structure of MnP. (b) and (c) Observed (F$_{obs}$) vs calculated (F$_{calc}$) structure factors of SCXRD data at $T$ = 295 K and $T$ = 30 K respectively. (d) XRD pattern of single crystalline MnP showing $(h~0~l)$ diffraction peaks. (e) Enlarged HRTEM image displaying $(1~0~2)$ plane. (f) SAED pattern displaying different planes. }
\label{fig:XRD}
\end{figure*}


\section{Results}
\subsection{Sample Characterizations}
The present study on MnP was carried out on single crystal sample prepared using Sn flux [see Appendix A]. Room-temperature single crystal x-ray diffraction (SCXRD) refinement confirms that MnP crystallizes in an orthorhombic structure [Fig.~\ref{fig:XRD}(a)] with space group $Pnma$ (No. 62; $c>a>b$). Low-temperature SCXRD data collected at 30 K reveal no structural phase transition, confirming the stability of the orthorhombic $Pnma$ structure. The good linear correlation between the observed ($F_{obs}$) and calculated ($F_{cal}$) structure factors demonstrates the high quality of the refinement [Fig.~\ref{fig:XRD}(b), (c)]. The refined lattice parameters and other refinement details are listed in Table~\ref{table:xrd}. X-ray diffraction measurements taken from the flat surface of the crystal [as shown in the inset of Fig.~\ref{fig:XRD} (d)] show only ($h$ 0 $l$) Bragg reflections, indicating that the flat surface is parallel to the ($h$ 0 $l$) plane and the crystal is grown along the $b$-axis. The HRTEM image shown in Fig.~\ref{fig:XRD} (e) further confirms the undistorted plane or the absence of disorder in the crystal. The inset of Fig.~\ref{fig:XRD} (e) shows that the spacing between the plane is 2.57 \AA, which is well in agreement with the interplanar spacing $d_{(102)}$ of (1 0 2) planes as obtained from the x-ray diffraction studies. Fig.~\ref{fig:XRD} (f) shows the Selected Area Electron Diffraction (SAED) pattern with the assigned Miller indices.

\begin{table}
    \caption{Refined structural parameters of our MnP sample obtain from SCXRD at $T$=295 K and $T$=30 K.}
	\label{table:xrd}
	\centering
    \setlength{\tabcolsep}{7pt}
	\begin{tabular}{c   c   c   c   c  c}
		\hline
        \hline
		Method &  & SCXRD &  & SCXRD &   \\
        Temperature &  &295 K  & &30 K  &   \\
        Chemical Formula & &MnP  &  &MnP & \\
        Space Group &  & \textit{Pnma}  &  & \textit{Pnma} &  \\
        a(\AA)  &  & 5.2534(4)  & & 5.2326(2) &  \\
        b(\AA)  &  & 3.16986(16)  &  & 3.18120(10) &  \\
        c(\AA)  &  & 5.9153(4)  &  & 5.8933(4) &  \\
        V(\AA$^3$) &  & 98.505(11) &   & 98.100(8) &  \\
        $R$, $wR2$ &  & 2.56, 5.70 &   & 2.85, 6.48  &  \\
        $GOF$ &  & 2.05 &   & 2.45 &  \\
        \hline
	\end{tabular}
    \setlength{\tabcolsep}{3.7pt}
    \begin{tabular}{c   c   c   c   c  c}    
        Atom &  Wyckoff  & x & y & z & Occupancy\\
        \hline
        Mn & 4c & 0.00474(8) & 0.25 & 0.19661(7) & 1.0 \\
        P &  4c & 0.18795(14) & 0.25 & 0.56852(11) & 1.0 \\
        \hline
        \hline
    \end{tabular}
\end{table}

\subsection{Magnetization}

\begin{figure*}
\centering
\includegraphics[width = 15 cm]{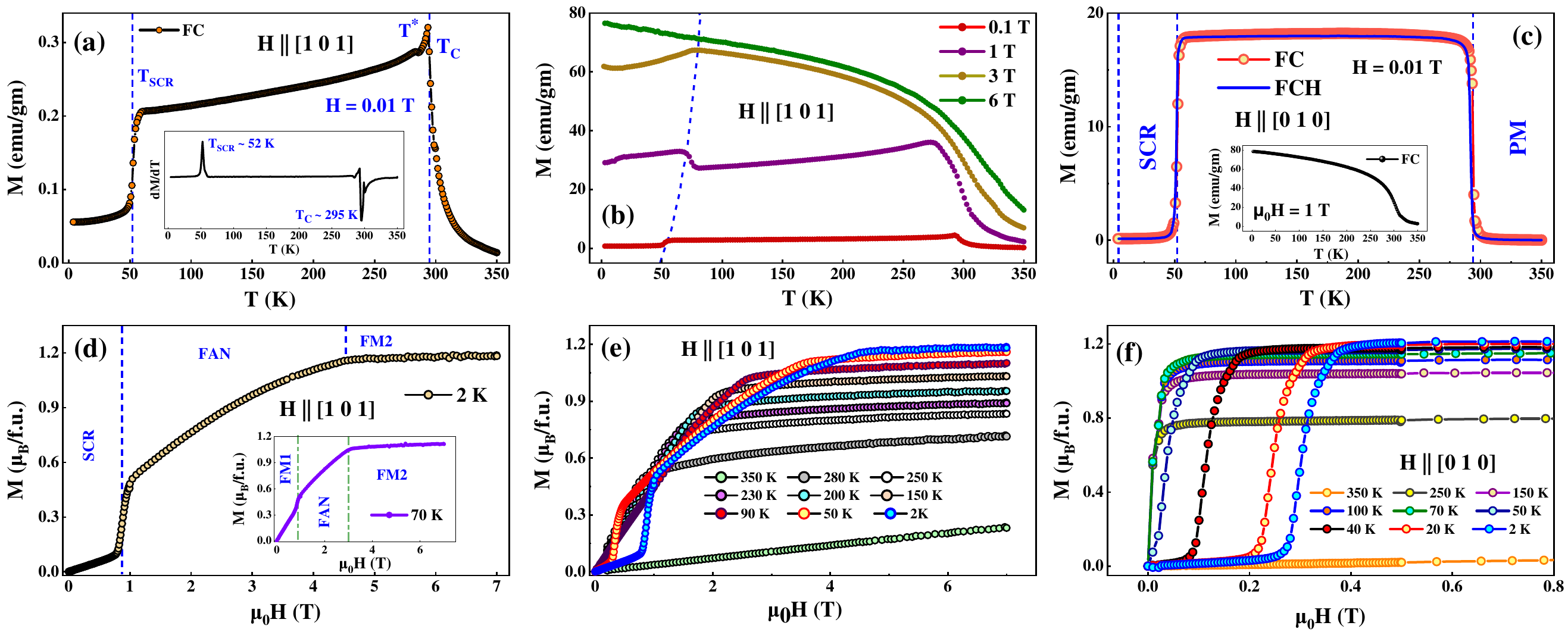}
\caption{\textbf{Magnetic properties of MnP.} (a) Temperature ($T$) dependence of magnetization ($M$) in field-cooled (FC) protocols under  $\mu_0H$= 0.01 T. The inset shows the temperature derivative of $M(T)$ for $\mu_0H\parallel$ [1 0 1] direction. (b) $M$ vs. $T$ data under different fields ($\mu_0H$=0.1 T, 1 T, 3 T, 6 T) for $\mu_0H\parallel$ [1 0 1] direction. (c) $M$ vs. $T$ data in field-cooled (FC) and field-cooled-heating (FCH) protocol under $\mu_0H$= 0.01 T for $\mu_0H\parallel$ [0 1 0] direction and the inset shows the FC curve under $\mu_0H$= 1 T. (d) $M$ vs. $\mu_0H$ curve at 2 K for $\mu_0H\parallel$ [1 0 1] direction and the inset shows $M$ vs. $\mu_0H$ curve at 70 K. (e) $M$ vs. $\mu_0H$ curve at different constant $T$ with $\mu_0H\parallel$ [1 0 1] direction. (f) $M$ vs. $\mu_0H$ curve at different constant $T$ with $\mu_0H\parallel$ [0 1 0] direction.} 
\label{fig:MTMH}
\end{figure*}

To verify the complex and highly anisotropic magnetic phase diagram of MnP [Fig.~\ref{fig:Phase}(a)-(c)], temperature ($T$) and magnetic field ($H$) dependent magnetization ($M$) measurements were performed. Figure~\ref{fig:MTMH}(a) shows the field-cooled (FC) $M(T)$ data measured at $\mu_0H=$0.01 T for $H\parallel[1~0~1]$, exhibiting a paramagnetic (PM) to ferromagnetic (FM1) transition at $T_C\sim$295 K. A small kink at $T^*$ just below $T_C$ is attributed to a slight canting of Mn spins toward the $a$-axis~\cite{MnP_JPSJ_Neutron,T_star1,T_star2}. On further cooling, $M$ decreases sharply at $T_{\mathrm{SCR}}\sim$52 K, marking the first-order FM1-to-SCR transition~\cite{MnP_1980,MnP_1966}. Figure~\ref{fig:MTMH}(b) shows the FC $M(T)$ data measured under different magnetic fields ($\mu_0H =$ 0.1, 1, 3, and 6 T), revealing distinct changes near $T_{\mathrm{SCR}}\sim$52 K as the system traverses different magnetic phases. For $H\parallel[0~1~0]$ [Fig.~\ref{fig:MTMH}(c)], the sample exhibits the same PM-FM1 and FM1-SCR transitions, but with a substantially larger magnetization than for $H\parallel[1~0~1]$, highlighting the strong magnetic anisotropy of MnP. The inset of Fig.~\ref{fig:MTMH}(c) shows that the SCR transition is suppressed at $\mu_0H=$1 T, consistent with the magnetic phase diagram [Fig.~\ref{fig:Phase}(c)].
 
\par
To further explore the magnetic phase diagram of MnP, isothermal $M(\mu_0H)$ measurements were performed at various temperatures for $\mu_0H\parallel[1~0~1]$ and $\mu_0H\parallel[0~1~0]$. For $\mu_0H\parallel[1~0~1]$, the 2 K $M(\mu_0H)$ curve [Fig.~\ref{fig:MTMH}(d)] exhibits successive SCR-FAN and FAN-FM2 transitions at $\mu_0H\approx$0.85 T and $\mu_0H>$4 T, respectively, with a saturation magnetization of $\sim$1.18~$\mu_B$/f.u. As $T$ increases, the SCR-FAN transition shifts to lower fields [Fig.~\ref{fig:MTMH}(e)]. In the range 50$<T<$110 K, the system undergoes successive FM1-FAN-FM2 transitions, while only an FM1-FM2 transition remains above 110 K. In contrast, for $\mu_0H\parallel[0~1~0]$, the FAN phase is absent and only an SCR-to-ferromagnetic transition is observed below 50 K. This transition occurs near 0.3 T at 2 K and shifts to lower fields with increasing temperature [Fig.~\ref{fig:MTMH}(f)]. Above 50~K, no distinct field-induced transition is observed, with $M$ rapidly approaching saturation. The saturation magnetization reaches $\sim1.2~\mu_B$/f.u. at 2~K, comparable to that for $\mu_0H\parallel[1~0~1]$.

\subsection{XMCD}
\begin{figure}[htbp!]
	\centering
	\includegraphics[width = 8.2 cm]{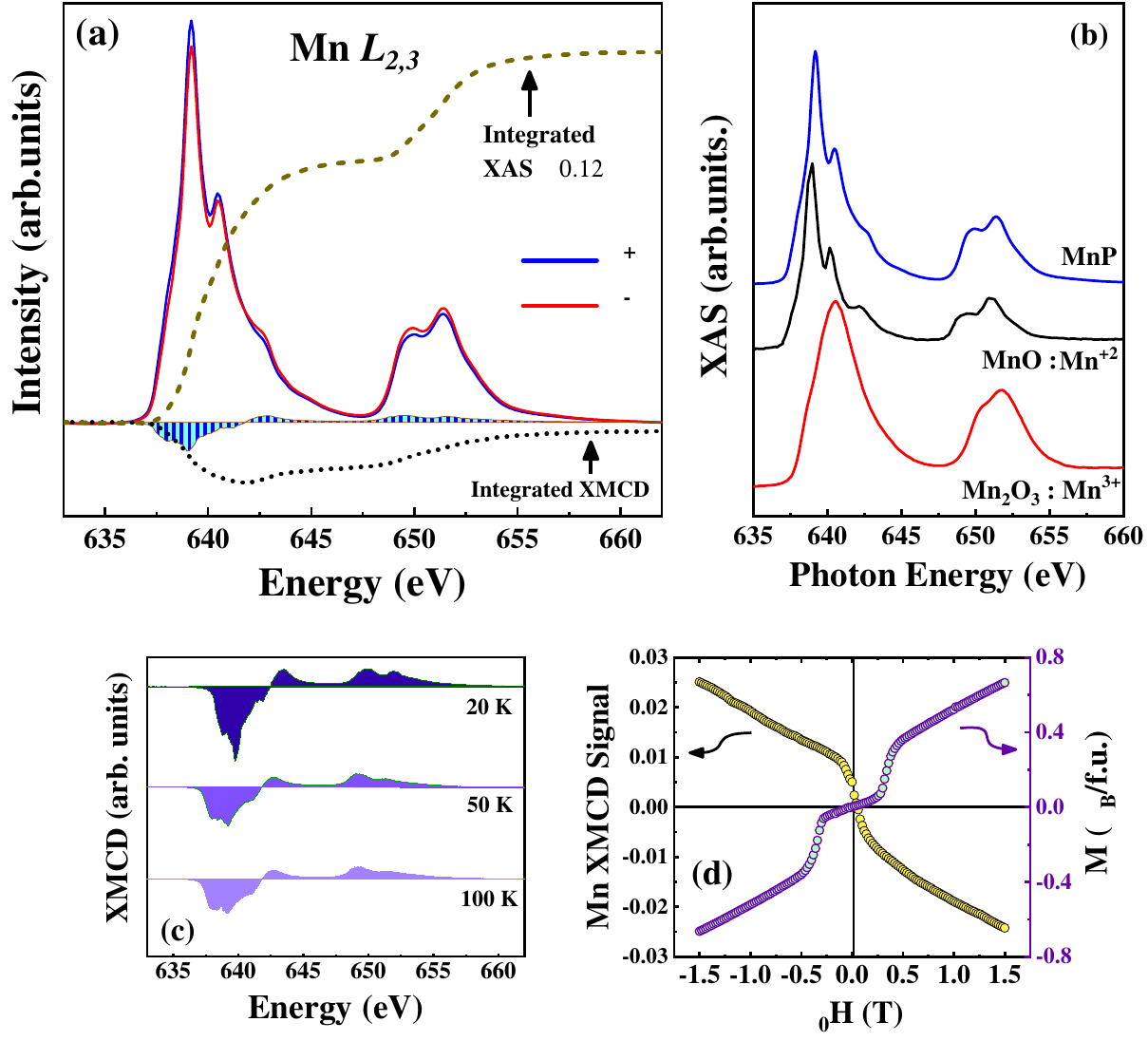}
	\caption{\textbf{XAS and XMCD of MnP.} (a) Mn $L_{2,3}$-edge XAS and XMCD spectra of MnP measured at 20 K with $\mu_0H \parallel$ [1 0 1]. (b) Comparison of the Mn $L_{2,3}$-edge XAS spectra of MnP, MnO, and Mn$_2$O$_3$. (c) Mn $L_{2,3}$-edge XMCD spectra measured at 20, 50, and 100 K. (d) Comparison of the field dependence of the Mn $L_3$-edge XMCD intensity (left panel) and the bulk magnetization (right panel) measured at 50 K.}
	\label{fig:XMCD}
\end{figure}

Figure~\ref{fig:XMCD}(a) presents the Mn $L_{2,3}$-edge x-ray absorption spectroscopy (XAS) and x-ray magnetic circular dichroism (XMCD) spectra of MnP measured at $T =$ 20 K under an applied magnetic field of 1.5 T ($\mu_0H \parallel$ [1 0 1]). Here, $\mu^{+}$ and $\mu^{-}$ denote the XAS intensities measured with opposite photon helicities relative to the sample magnetization direction, while the XMCD spectrum corresponds to their difference ($\mu^{-}-\mu^{+}$).
\par
As shown in Fig.~\ref{fig:XMCD}(b), the measured Mn $L_{2,3}$-edge XAS spectrum of MnP closely resembles that of MnO (Mn$^{2+}$), while it differs significantly from that of Mn$_2$O$_3$ (Mn$^{3+}$). This similarity suggests that the Mn ions in MnP are predominantly in a nearly divalent state. The subtle differences in the spectral line shape between MnP and MnO can be attributed to their distinct local crystal environments and symmetries.
\par
The XAS and XMCD measurements were also repeated at elevated temperatures ($T=$ 50, 100 K) and dichroism was also observed [see Fig.~\ref{fig:XMCD}(c)]. We have calculated the spin ($\mu_s$) and orbital ($\mu_l$) magnetic moments at all these $T$s using the sum rules~\cite{XMCD_rule1,XMCD_rule2} and the results are listed in the table~\ref{tab:moment}.

\begin{equation}
\resizebox{0.95\linewidth}{!}{$
\begin{aligned}
\mu_s = &-n_h
\frac{
6\int_{L_3}(\mu^--\mu^+)dE
-4\int_{L_{2,3}}(\mu^--\mu^+)dE
}{
\int_{L_{2,3}}(\mu^-+\mu^+)dE
}
\times SC \\
&-\langle T \rangle
\end{aligned}
$}
\end{equation}

\begin{equation}
    \mu_l=-\frac{4}{3}n_h\frac{\int_{L_{2,3}}(\mu^--\mu^+)dE}{\int_{L_{2,3}}(\mu^-+\mu^+)dE}
\end{equation}

\noindent where $n_h$, SC and $\langle T \rangle$ are the number of $d$ holes, spin correction factor (estimated to be 1.5 for Mn~\cite{XMCD_Mn2VAl,XMCD_Mn_Film}) and the averaged magnetic dipole term (can be neglected due to its rather small contribution~\cite{XMCD_Mn3Sn}), respectively. The obtained total magnetic moment ($\mu_{\mathrm{total}}$) from the XMCD analysis is $0.59~\mu_B$/Mn at 20 K ($\mu_s=$0.57$~\mu_B$/Mn and $\mu_l=$0.02$~\mu_B$/Mn), which is in good agreement with the bulk magnetization value of $0.63~\mu_B$/Mn measured at the same temperature. As shown in Fig.~\ref{fig:XMCD}(d), the field dependence of the Mn XMCD signal at the $L_3$ edge closely matches that of the bulk magnetization measured at 50 K, confirming that the XMCD response faithfully reflects the magnetic behavior of MnP. It is to be noted that Mn possesses a small but finite orbital moment of 0.02$~\mu_B$/Mn, despite the fact that Mn 3$d$ state is half filled with five electrons. 

\begin{table}[htbp]
\centering
\caption{Spin, orbital, and total magnetic moments obtained at different temperatures.}
\label{tab:moment}
\begin{tabular}{cccc}
\hline\hline
Temperature &
$\mu_s$ &
$\mu_l$ &
$\mu_{\mathrm{total}}$ \\
(K) &
($\mu_B$/atom) &
($\mu_B$/atom) &
($\mu_B$/atom) \\
\hline

20  & $0.57$ & $0.02$ & $0.59$ \\
50  & $0.51$ & $0.04$ & $0.55$ \\
100 & $0.50 $ & $0.04$ & $0.54$ \\

\hline\hline
\end{tabular}
\end{table}

\subsection{Electrical Resistivity \& Magnetoresistance}
\par

\begin{figure}[ht]
	\centering
	\includegraphics[width = 7 cm]{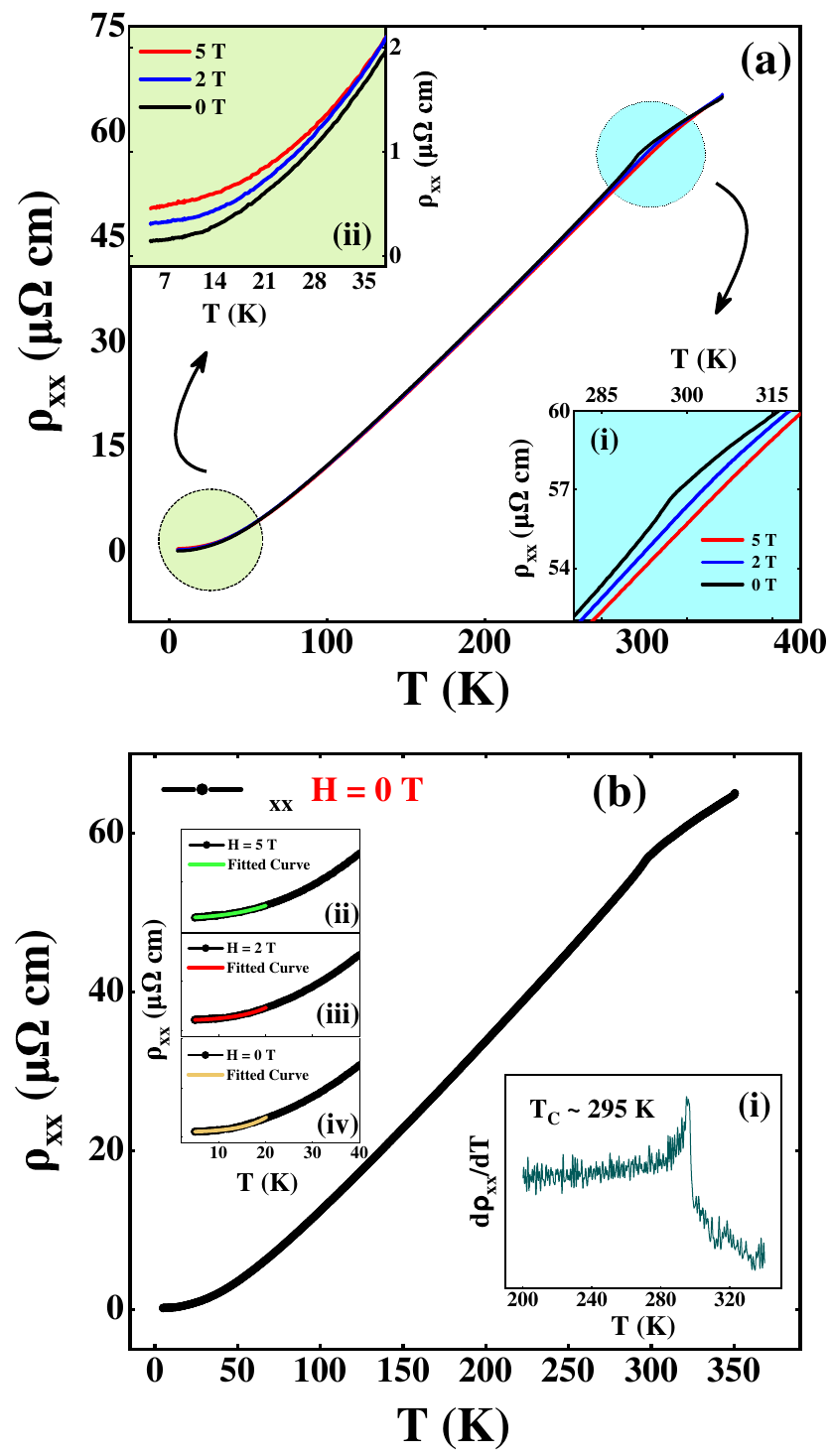}
	\caption{\textbf{Electrical transport properties of MnP.} (a) Shows temperature ($T$) dependence of longitudinal resistivity ($\rho_{xx}$) for different applied magnetic field, $\mu_0H= 0, 2, 5$ T, along [1 0 1] direction. The inset a(i) and a(ii) shows the field dependent $\rho_{xx}$ vs $T$ data at high temperature and low temperature respectively. (b) Shows zero field $\rho_{xx}$ vs $T$ data with the inset (i) showing $\frac{d\rho_{xx}}{dT}$ vs $T$ curve. The fitting of low temperature $\rho_{xx}$ data at different field has been shown in inset b(ii), b(iii) and b(iv).}
	\label{fig:RT}
\end{figure}

Despite having a complex magnetic phase diagram, the magneto-transport behavior of MnP remains less explored. To investigate this, we first measured the $T$ dependence of the longitudinal resistivity ($\rho_{xx}$) at different applied fields ($\mu_0H$ = 0, 2T, 5T), with the current ($I$) applied along the [0 1 0] direction with $\mu_0H\parallel$[1 0 1] direction [the plots are shown in Fig.~\ref{fig:RT} (a)]. The zero-field $\rho_{xx}$ vs. $T$ shows good metallic nature, with a kink at the ferromagnetic transition temperature ($T_C \sim 295$ K), determined from the $d\rho_{xx}/dT$ vs. $T$ graph [see inset (i) of Fig.~\ref{fig:RT} (b)]. The studied sample has a residual resistivity of $\rho_0$ (5 K) = 0.139 $\mu\Omega$-cm and a residual resistivity ratio (RRR) of $\rho_{xx}(350 K)/\rho_{xx}(5 K) =$ 467, indicating a very high-quality crystal grown by the Sn-flux method. The inset (i) of Fig.~\ref{fig:RT}(a) shows that $\rho_{xx}$ decreases with increasing $H$ at high $T$, reflecting the suppression of spin-dependent $s-d$ scattering. In contrast, inset (ii) of Fig.~\ref{fig:RT}(a) demonstrates that for $T<T_{SCR}$, $\rho_{xx}$ increases with increasing field, indicating an enhancement of $s-d$ scattering and the emergence of positive MR.
 
\par
The $\rho_{xx}(T,H)$ data for MnP were found to fit well with the expression in the low-$T$ region ($T\leq$ 20 K):
\begin{equation}
    \rho_{xx}(T) = \rho_0 + \alpha_{mag}T^2 + \alpha_{ph}T^5
\label{eqn:RT}
\end{equation}

 Here $\rho_0$ represents $T$ independent residual resistivity, $\alpha_{mag}$ and $\alpha_{ph}$ represent contributions from electron-magnon and electron-phonon scattering respectively. The fitted curves are shown in the inset (ii), (iii) and (iv) of Fig.~\ref{fig:RT} (b) for different applied field and the corresponding coefficients are shown in table~\ref{table:RT}. Table~\ref{table:RT} shows that the $\alpha_{mag}$ increases systematically with increasing $H$, indicating an enhancement of the $s-d$ scattering as the system evolves through the field-induced complex magnetic phases (SCR and FAN). At zero field, the value of $\alpha_{mag}$ is slightly higher than that reported for its closely related compound CrP ($2.9\times10^{-4}$ $\mu\Omega$ cm K$^{-2}$)~\cite{CrP_MR}, which may be attributed to the more complex magnetic structure of MnP. In contrast, the value of the coefficient $\alpha_{ph}$ is very small and decreases with increasing $H$.

\begin{table}
\caption{The fitting results of the coefficients of Eqs.~\ref{eqn:RT} for all the applied fields 0, 2T, 5T.}
	\label{table:RT}
	\centering
	\begin{tabular}{c c c c c}
		\hline
        \hline
        \noalign{\vspace{2mm}}
		  & &$\rho_{xx}(T) = \rho_0 + \alpha_{mag}T^2 + \alpha_{ph}T^5$ \\
          \noalign{\vspace{2mm}}
          \hline
          $\mu_0H$ &   $\rho_0$  &  $\alpha_{mag}$($\times10^{-4}$)  &  $\alpha_{ph}$($\times10^{-8})$  \\
		(T) &   ($\mu\Omega$ cm)  &  ($\mu\Omega$ cm K$^{-2}$)  &  ($\mu\Omega$ cm K$^{-5}$)  \\
        \hline
		0 & 0.139    &  4.815   &   7.574  \\
		
		2 & 0.3    &  5.433   & 5.15     \\
	
            5 & 0.45    &  7.247   & 1.95     \\
            \hline
            \hline
	\end{tabular}
\end{table}

\par

\begin{figure}
\centering
\includegraphics[width=7cm]{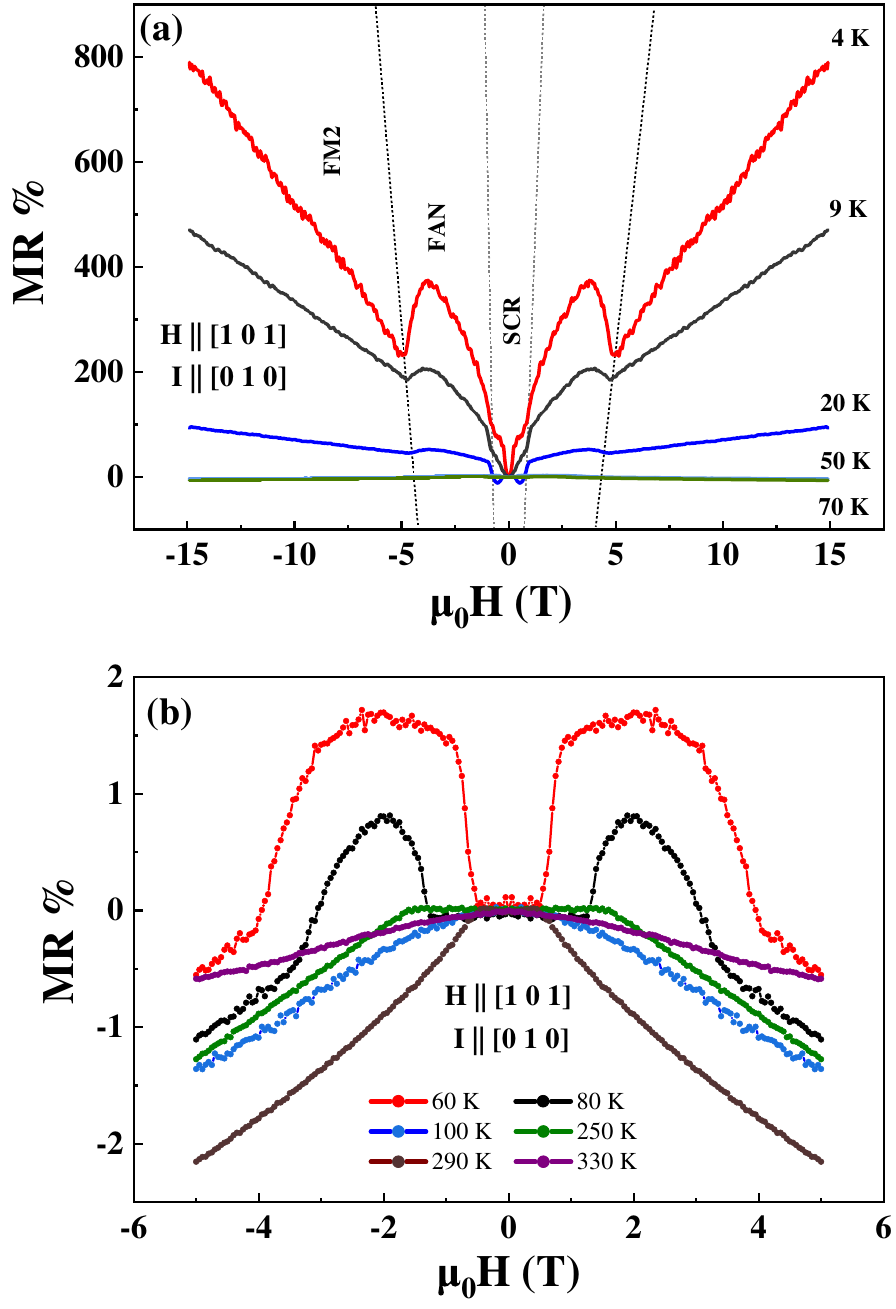}
\caption{\textbf{Magnetoresistance of MnP.} (a) Shows the MR \% vs $\mu_0H$ data at different constant $T$ measured between $\pm$15 T with different region marked. (b) Shows MR \% vs $\mu_0H$ data for $T>50$ K measured between $\pm$5 T.} 
\label{TMR}
\end{figure}

We also measured the field dependence of the MR at various constant $T$s. The MR is commonly calculated as the relative change in resistivity due to an applied magnetic field, defined by MR $ \%= [\rho_{xx}(\mu_0 H) - \rho_{xx}(0)]/\rho_{xx}(0) \times 100$ $\%$, where $\rho_{xx}(\mu_0 H)$ and $\rho_{xx}(0)$ are the resistivities under $\mu_0 H$ and at zero field, respectively. Figure~\ref{TMR} (a) and (b) show the transverse MR measured at different constant temperatures for $\mu_0 H \parallel$ [1 0 1] and $I \parallel$ [0 1 0] over the ranges $\pm 15$~T and $\pm 5$~T, respectively.
\par
For $T < 50$ K, the MR exhibits a positive, non-saturating field dependence up to the maximum applied field of $\mu_0H=\pm$15 T. In the low-field region, distinct anomalies associated with the SCR-to-FAN transition are observed, consistent with the magnetic phase diagram in Fig.~\ref{fig:Phase}(b). Upon increasing the field beyond $\mu_0H \gtrsim \pm$4 T, the system enters the FM2 phase, accompanied by a pronounced decrease in MR at the phase boundary due to the suppression of $s-d$ scattering as the spins align with the applied field. Beyond the transition, the MR exhibits an non-saturating linear field dependence throughout the FM2 phase. The MR reaches nearly $800\%$ at 4 K under $\mu_0H=\pm15$ T and drops substantially as $T$ increases.
\par
As shown in Fig.~\ref{TMR}(b), in the $T$ range 50 $\leq T \leq$ 90 K, the MR is weakly negative in the FM1 phase and gradually decreases with increasing temperature. At the FM1--FAN phase boundary, the MR exhibits a pronounced jump to positive values, remaining positive throughout the FAN phase, consistent with the magnetic phase diagram in Fig.~\ref{fig:Phase}(b). Upon entering the FM2 phase, the MR changes sign again and becomes negative, with its magnitude increasing monotonically with magnetic field. For $T \geq 100$ K, the MR remains weakly negative over the entire measured field range. We find that in the FAN phase (below 100 K), the MR is always positive, although the magnitude of MR diminishes as $T$ increases. In the FM1 phase above 100 K, the MR is found to be small negative. Such negative MR in the high $T$ FM1 phase can be attributed to the suppression of the $s-d$ scattering by the applied $H$. 
\par
At low $T$, the large non-saturating LPMR in the FM2 phase appears to have a quantum-mechanical origin associated with the underlying band topology of MnP, as discussed in detail later. Furthermore, the Kohler plot [MR\% \emph{vs.} $\mu_0H/\rho_{xx}(0)$] exhibits a clear violation of Kohler's rule (see Appendix D), suggesting contributions from more than one scattering processes~\cite{SmMn2Ge2}.

\subsection{Hall Resistivity Study}

\par
The Hall resistivity ($\rho_{xy}$) of MnP was measured as a function of $H$ at constant $T$s in the range 20$~\mathrm{K} \leq T \leq$ 330~$\mathrm{K}$ with $\mu_0H \parallel$ [1 0 1]. There are two different regimes in the Hall data reflecting distinct underlying mechanisms of the Hall effect: $T \geq 100$ K and $T < 100$ K. To eliminate the contribution from the longitudinal resistivity arising due to misalignment of the voltage leads, the measured $\rho_{xy}$ was anti-symmetrized using the expression $\rho_{xy}(\mu_0 H) = \frac{1}{2}[\rho_{xy}(+\mu_0 H) - \rho_{xy}(-\mu_0 H)]$.
\par
\begin{figure*}[ht]
\centering
\includegraphics[width = 16.5 cm]{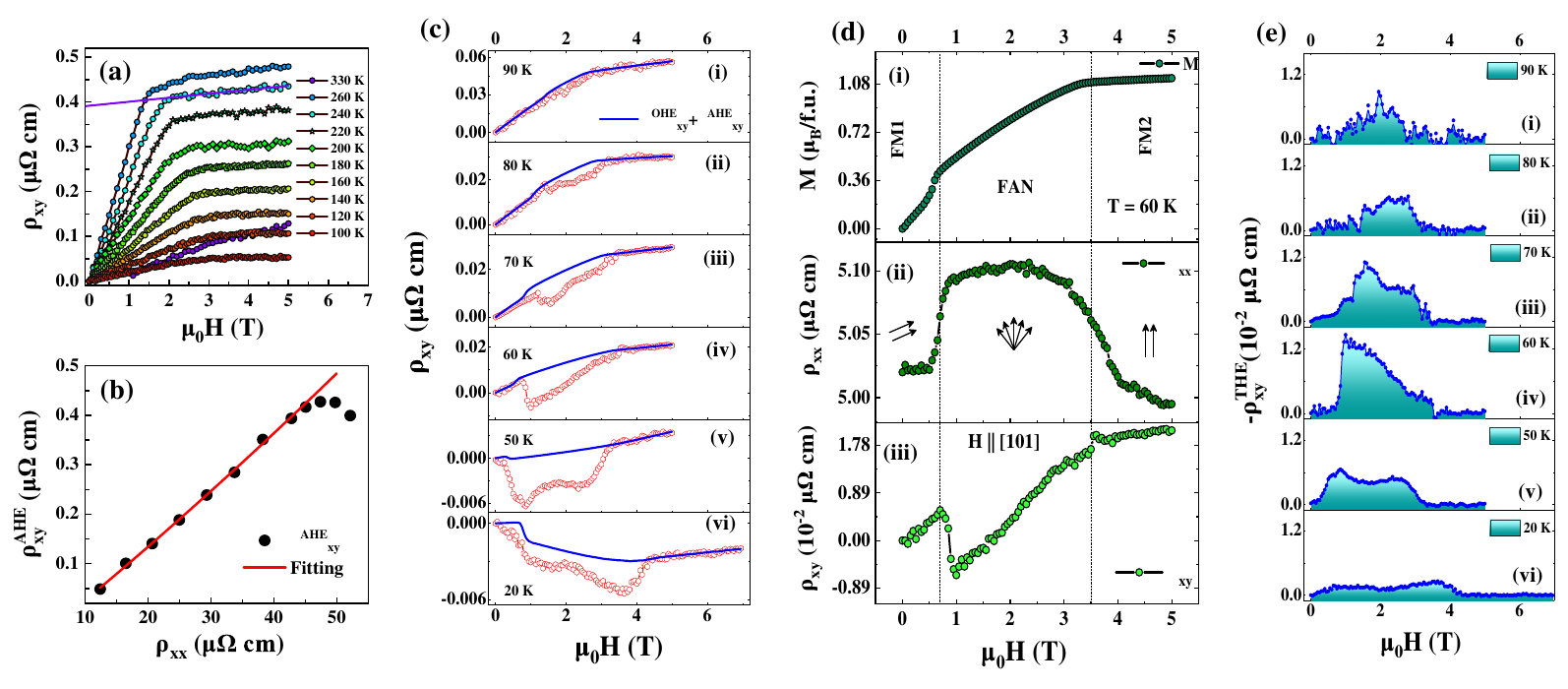}
\caption{\textbf{Anomalous and topological Hall effects in MnP.} (a) Hall resistivity $\rho_{xy}$ vs $\mu_0H$ at different constant $T$ for $\mu_{0}H \parallel [1~0~1]$. 
(b) Anomalous Hall resistivity $\rho_{xy}^{\mathrm{AHE}}$ as a function of $\rho_{xx}$, with the solid red line representing the fitting. 
(c) Field dependence of $\rho_{xy}$ at different $T$. The red open circles represent the experimental data, while the solid blue lines show the calculated Hall resistivity, $\rho_{xy}^{\mathrm{OHE}}+\rho_{xy}^{\mathrm{AHE}}$, obtained from the corresponding fitting parameters. 
(d) Field dependence of (i) $M$, (ii) $\rho_{xx}$, and (iii) $\rho_{xy}$ at $T=60$~K for $H \parallel [1~0~1]$. The vertical dashed lines indicate the characteristic field boundaries of the FM1, FAN, and FM2 magnetic phases. 
(e) Calculated topological Hall resistivity $-\rho_{xy}^{\mathrm{THE}}$ at different $T$, obtained after subtracting the ordinary and anomalous Hall contributions from the measured Hall resistivity.}
\label{fig:THE}
\end{figure*}

{\bfseries\boldmath $T \geq 100$ K:} Below $T_C$, $\rho_{xy}$ exhibits a non-linear dependence on $\mu_0 H$, showing a sharp increase in the low-field region with a tendency to saturate at higher fields, indicating the presence of AHE in the system [Fig.~\ref{fig:THE}(a)]. In this region $\rho_{xy}$ of MnP can be expressed by an empirical formula~\cite{Fe3Sn2,Cr2.76Te4},

\begin{equation}
    \rho_{xy}=\rho_{xy}^{OHE}+\rho_{xy}^{AHE}=R_0\mu_0H+R_s\mu_0M,
    \label{AHE}
\end{equation}

\noindent Here, $\rho_{xy}^{OHE}$ is the Lorentz force-induced ordinary Hall resistivity, and $\rho_{xy}^{AHE}$ is the anomalous Hall resistivity. The coefficients of the ordinary and anomalous Hall resistivities, $R_0$ and $R_s$, can be obtained by linearly extrapolating the high-field $\rho_{xy}(\mu_0 H)$ data to $\mu_0 H = 0$~\cite{Fe3Sn2}. The carrier concentration ($n_h$), calculated from $R_0$ using $R_0 = \frac{1}{n_h e}$, is found to be positive over the entire measured temperature range, indicating that holes are the majority charge carriers.

\par
The anomalous component of $\rho_{xy}$ depends on both the intrinsic and extrinsic properties of the material, and scaling analysis is essential to understand the underlying mechanisms. The AHE is primarily governed by three mechanisms, which follow the scaling relation $\rho_{xy}^{AHE} \propto \rho_{xx}^q$, where $q = 1$ corresponds to skew scattering, and $q = 2$ corresponds to intrinsic and side-jump contributions. We use the following scaling relation~\cite{CrTe_PRM}:

\begin{equation}
    \rho_{xy}^{AHE}=\alpha'\rho_{0}+\alpha''\rho_{xxT}+b\rho_{xx}^2,
    \label{TYJ_original}
\end{equation}

\noindent The parameters $\alpha'$ and $\alpha''$ correspond to skew scattering induced by impurities and phonons, respectively. Here, $\rho_{xxT}(=\rho_{xx} - \rho_{0})$ represents the resistivity due to dynamic phonon scattering. The corresponding fit is shown in Fig.~\ref{fig:THE}(b). The value of $\rho_{0}$ was kept constant at 0.139~$\mu\Omega$ cm, and the fitting yields the parameters $\alpha' = -0.552$, $\alpha'' = 0.0102$, and $b = 2.108 \times 10^{-5}(\mu\Omega)^{-1}$cm$^{-1}$ (= 21.08 S cm$^{-1}$). The scaling analysis indicates that the AHE in MnP is primarily dominated by skew scattering, but also includes a very small contributions from the side-jump mechanism and the Berry-curvature-induced intrinsic mechanism in the temperature range 100  $\leq T \leq$330 K. Since the present MnP crystal is weakly disordered (RRR = 467), the dominance of skew scattering mechanism toward AHE is expected~\cite{Nagaosa1, Nagaosa2}.
\par

\par

\subsection{Topological Hall Effect}

\par
{\bfseries\boldmath $T < 100$ K:} As we go below $T = 100$ K, $\rho_{xy}$ begins to develop additional curvature in the FAN phase and also in the SCR phase for $T \leq 50$ K, as shown in Fig.~\ref{fig:THE}(c). By comparing the variation of $M$, $\rho_{xx}$, and $\rho_{xy}$ with $\mu_0 H$ below 100 K [see Fig.~\ref{fig:THE}(d) for $T = 60$ K], it is clear that Eqn.~\ref{AHE} alone is insufficient to describe the behavior of $\rho_{xy}$ in this temperature range, since the curvature of $\rho_{xy}$ is opposite to that of $M$ and $\rho_{xx}$. This indicates the presence of an additional contribution to the AHE in the FAN and SCR phases. To account for this contribution in MnP, we employ the following expression~\cite{THE_Thamizhavel}:

\begin{equation}
    \rho_{xy}=R_0\mu_0H+S_H\rho_{xx}^2M(1+\alpha\frac{\rho_0}{\rho_{xx}})+\rho_{xy}^{THE},
    \label{THE}
\end{equation}

\noindent Here, $S_H \rho_{xx}^2$ (equivalent to $\mu_0 R_s$) corresponds to the AHE coefficient, $\rho_0$ represents the zero-field resistivity at a given temperature, and $\rho_{xy}^{THE}$ is the topological Hall resistivity. Since our previous scaling analysis indicates that the AHE in MnP includes contributions from both skew scattering and side-jump/intrinsic mechanisms, we have included both terms in Eqn.~\ref{THE}. The blue lines in Fig.~\ref{fig:THE}(b)(i)–(vi) represent the fitted $\rho_{xy}^{OHE}+\rho_{xy}^{AHE}$ contribution. The topological Hall resistivity, $\rho_{xy}^{THE}$, is subsequently obtained by subtracting the fitted $\rho_{xy}^{OHE}+\rho_{xy}^{AHE}$ component from the measured Hall resistivity, $\rho_{xy}$. The extracted THE components at different temperatures are plotted in Fig.~\ref{fig:THE} (e). From Fig.~\ref{fig:THE} (e), it is evident that the THE signal appears only in the FAN phase for 50K$< T <$100K, and both in SCR and FAN for T$\leq$50 K, as shown in Fig.~\ref{fig:THE}(e) (v) and (vi) for 50 K and 20 K, respectively.

\par
Shiomi \emph{et al.}~\cite{MnP_THE} previously reported the presence of a THE in the FAN phase of MnP and argued that the apparent THE observed in the SCR phase might be an artifact, assuming the net scalar spin chirality in this phase to be zero. Later, Yamazaki \emph{et al.}~\cite{MnP_JPSJ_Neutron} conducted a detailed single-crystal neutron diffraction study on MnP and showed that the helical planes in the SCR phase are tilted towards the $c$-axis from the $ab$-plane by angles $\theta$ and $-\theta$, alternating along the $a$-axis. Their results suggest that the scalar spin chirality in the SCR phase may, in fact, be non-zero, giving rise to a finite contribution to the THE in this phase. In our case we applied the magnetic field along [1 0 1] direction and according to the phase diagram for $H\parallel$[0 0 1], there is a CONE phase in the observed THE region [Fig.~\ref{fig:Phase} (a)]. So there might be a combined effect from tilted helical phase and conical phase which leads to the appearance of THE below the FAN phase.

\begin{figure*}
\centering
\includegraphics[width = 17 cm]{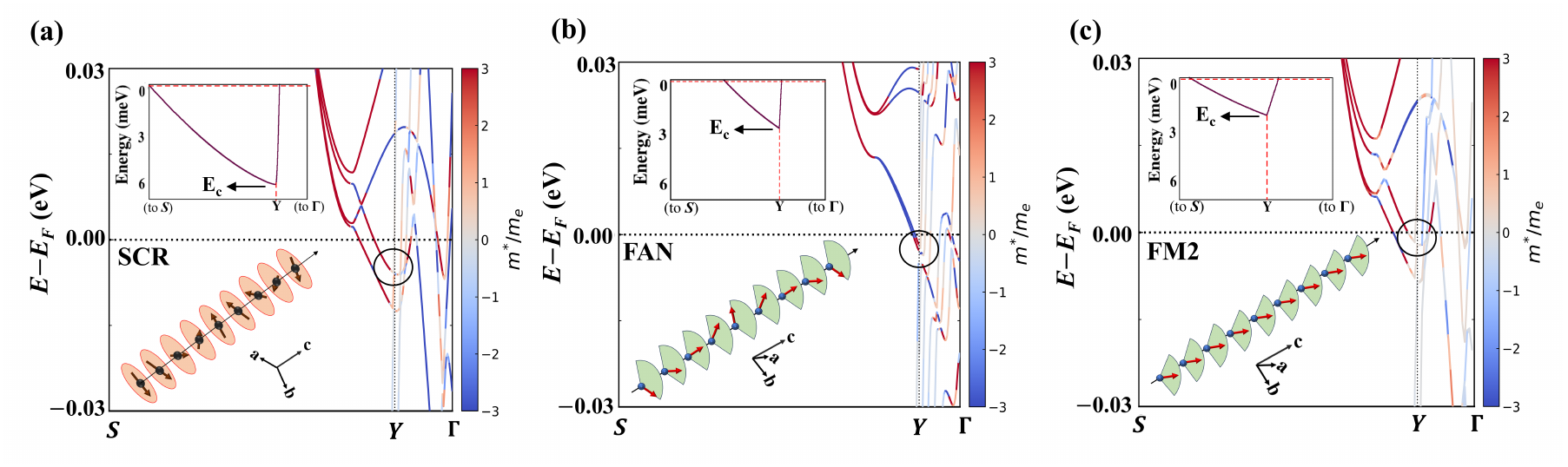}
\caption{ \textbf{Magnetic-state evolution of the semi-Dirac-like electronic structure in MnP.} Calculated low-energy band structures along the $S-Y-\Gamma$ path of the orthorhombic Brillouin zone for the (a) SCR, (b) FAN, and (c) FM2 states. The high-symmetry-point notation follows the convention of Bradley and Cracknell~\cite{BradleyCracknell}. The lower insets schematically illustrate the corresponding magnetic configurations, while the upper insets magnify the low-energy band dispersion in the vicinity of the $Y$ point. The bands are colored according to the signed local  mass, $m^{*}/m_e$, where positive and negative values denote electron-like and hole-like curvature, respectively. A strongly anisotropic semi-Dirac-like crossing is observed near $Y$, with linear and quadratic dispersions along the two momentum directions adjoining the crossing. The crossing progressively approaches the $E_F$ upon evolution from SCR to FAN and FM2, with $E_F-E_c\simeq6.0$, $2.8$, and $2.3$~meV, respectively. The enhanced negative-curvature contribution near $E_F$ indicates an increasing hole-like character of the low-energy electronic states in the field-induced phases. Details of the directional effective-mass evaluation and the low-energy band fitting are provided in the Appendix C. }
\label{fig:DFT}
\end{figure*}

\subsection{First-principles electronic structure and semi-Dirac dispersion}

To understand the microscopic origin of the low-temperature magnetotransport in MnP, we investigated the evolution of the low-energy electronic structure across the SCR, FAN, and FM2 magnetic phases. Figure~\ref{fig:DFT} shows the calculated band structures along the $S-Y-\Gamma$ path of the orthorhombic Brillouin zone for the three magnetic configurations. The high-symmetry-point notation follows the convention of Bradley and Cracknell~\cite{BradleyCracknell}. In all three phases, a strongly anisotropic band crossing is observed in the vicinity of the $Y$ point and close to the $E_F$. Most importantly, the dispersion exhibits qualitatively distinct momentum dependence on the two sides of the $Y$ point. Along $Y$--$\Gamma$, the band displays an approximately linear dispersion, whereas along $S$--$Y$, the corresponding branch exhibits pronounced quadratic curvature. We denote the energy at which these linear and quadratic branches meet at the $Y$ point as $E_c$.
 The coexistence of linear and quadratic dispersions around the same low-energy crossing constitutes the characteristic signature of a semi-Dirac-like electronic state~\cite{CrAs, FeP}.

\par
A particularly important feature is the strong dependence of the crossing energy on the magnetic configuration. In the SCR phase, the crossing is located approximately $6.0$~meV below the $E_F$. Upon transformation to the FAN state, the crossing shifts toward $E_F$, reducing the energy separation to approximately $2.8$~meV. In the field-polarized FM2 phase, it moves even closer to the $E_F$, with $E_F-E_c \simeq 2.3$~meV. Therefore, the field-induced evolution of the magnetic state provides an effective means of tuning the semi-Dirac-like electronic structure without removing its strongly anisotropic dispersion.
\par
The bands in Figure~\ref{fig:DFT} are additionally represented according to the effective  mass, $m^{*}$~\cite{suzuki1995first}. The effective mass is defined as \begin{equation} m^{*} = \frac{\hbar^{2}} {\displaystyle \frac{\partial^{2}E} {\partial k^{2}}}, \end{equation} where $k$ denotes the momentum coordinate along the corresponding reciprocal-space direction. Positive and negative values of $m^{*}$ correspond to electron-like and hole-like band curvature, respectively. Details of the numerical evaluation of $m^{*}$, together with the linear and quadratic fitting procedure used to extract the characteristic band velocity and effective mass, are provided in the Appendix C.

With the evolution from the SCR state to FAN and subsequently to FM2, an increasing contribution
from negative-curvature bands appears at the $E_F$. This indicates a progressive enhancement of hole-like low-energy states
in the field-induced phases. The result is consistent with the positive ordinary Hall coefficient obtained experimentally, which identifies holes as the majority charge carriers in MnP.

\subsection{Magnetic-state dependence of the low-energy parameters}
Quantitative fitting of the low-energy bands demonstrates that the characteristic parameters of the semi-Dirac-like dispersion remain of comparable magnitude across the SCR, FAN, and FM2 phases, while the position of the crossing relative to the Fermi energy evolves substantially. The quadratic branch yields $|m^{*}|/m_e = 0.68$, $0.70$, and $0.62$ for the SCR, FAN, and FM2 states, respectively. Here, $|m^{*}|$ denotes the magnitude of the  mass extracted from the quadratic branch in the vicinity of the semi-Dirac-like crossing. The corresponding velocities extracted from the linear branch are $8.9\times10^{4}$, $8.2\times10^{4}$, and $7.8\times10^{4}$~m~s$^{-1}$, respectively. The fitting procedure and associated uncertainties are provided in the Appendix C.

The comparatively modest changes in $v$ and $m^{*}$ indicate that the essential anisotropic dispersion survives the magnetic-state evolution. The dominant change instead occurs in the nodal energy: $E_{F}-E_{c}$ decreases from approximately $6.0$~meV in SCR to $2.8$~meV in FAN and $2.3$~meV in FM2. Thus, the principal role of the field-induced magnetic reconstruction is to bring the semi-Dirac-like band progressively closer to the chemical potential.

As discussed in detail in the Discussion section, this evolution has a direct consequence for high-field transport, particularly in the FM2 phase. As $E_{F}-E_{c}$ decreases, the characteristic magnetic field required to drive the associated small Fermi pocket into the quantum-limit regime is substantially reduced. The magnetic transition therefore modifies not only the spin-dependent scattering environment but also the energy scale governing Landau quantization of the low-energy carriers.

\begin{table*}[t]
\centering
\caption{ Effective mass $|m^{*}|/m_e$ extracted from the quadratic branch, characteristic band velocity $|v|$ extracted from the linear branch, energy separation $E_F-E_c$ of the semi-Dirac-like crossing, and estimated critical magnetic fields $\mu_{0}H^c$ for the SCR, FAN, and FM2 phases.}
\label{tab:effective_mass_velocity}
\begin{tabular}{lccccc}
\hline
\hline
Magnetic phase ~& ~$|m^{*}|/m_e$ ~& ~$|v|$ (m\,s$^{-1}$) ~&
$E_F-E_c$ (meV) ~& ~$\mu_0H^c(0)$ (T) ~&
$~\mu_0H^c(4\,\mathrm{K})$ (T)~\\
\hline
SCR & $0.68$ & $8.9\times10^4$ & $6.0$ & $9.89$ & $10.76$\\
FAN & $0.70$ & $8.2\times10^4$ & $2.8$ & $3.47$ & $4.13$\\
FM2 & $0.62$ & $7.8\times10^4$ & $2.3$ & $2.56$ & $3.15$\\
\hline
\hline
\end{tabular}
\end{table*}

\section{DISCUSSIONS}

The most significant finding of the present study is the observation of a large, non-saturating LPMR at low temperatures. Despite numerous proposed mechanisms, the origin of LPMR remains an open question. Three principal mechanisms have been widely proposed to explain LPMR: (i) the classical Parish-Littlewood (PL) model~\cite{PL_model}, (ii) electron-hole compensation, and (iii) Abrikosov's quantum linear magnetoresistance model~\cite{Abrikosov_2000,Abrikosov_1998}.
\par
The classical PL model attributes linear MR to spatial fluctuations in carrier mobility arising from inhomogeneities within the sample~\cite{PL_model}. In this framework, the disorder-induced mobility fluctuations lead to an admixture of the Hall voltage into the longitudinal resistivity, giving rise to a non-saturating linear MR. The effect is particularly pronounced in systems with large mobility variations, such as polycrystalline or highly disordered materials, such as Ag$_{2\pm\delta}$Se and Ag$_{2\pm\delta}$Te~\cite{PL_model}. However, this model is unlikely to account for the LPMR observed in MnP, as our sample exhibits an exceptionally low residual resistivity of 0.139~$\mu\Omega$-cm and a high RRR of 467, indicative of an excellent single-crystalline quality with minimal disorder.
\par
Another possible explanation for linear MR is electron-hole compensation arising from nearly balanced electron and hole carriers, as reported in ferromagnetic MnBi~\cite{MnBi}. Within the two-carrier model, this mechanism generally produces a quadratic (parabolic) field dependence of MR, as observed in materials such as WTe$_2$~\cite{WTe2} and CrP~\cite{CrP_MR}. However, this scenario is unlikely to apply to the present system, as our Hall effect analysis reveals a hole-dominated Hall response, providing no clear evidence for the nearly compensated electron-hole transport required to explain the observed MR within a compensation scenario.
\par
An alternative explanation is provided by the quantum linear MR model proposed by Abrikosov~\cite{Abrikosov_2000}, which predicts a non-saturating linear MR when a sufficiently strong magnetic field drives the system into the extreme quantum limit, confining all charge carriers to the lowest Landau level (LLL). Within this framework, the longitudinal resistivity and Hall coefficient are given by
\begin{equation}
\rho_{xx}=\rho_{yy}=\frac{N_i\mu_0H}{\pi n_0^2e}, \qquad
R=\frac{1}{n_0e},
\label{eq:Abrikosov_1}
\end{equation}
where $\rho_{ik}$ are the components of the resistivity tensor, $R$ is the Hall coefficient, $n_0$ is the carrier density, and $N_i$ is the concentration of scattering centers. The conditions for formulas~\ref{eq:Abrikosov_1} to apply are:
\begin{equation}
n_0 \ll \left(\frac{e\mu_0H}{\hbar }\right)^{3/2}, \qquad
k_BT \ll \frac{e\mu_0H\hbar}{m^{*}}.
\label{Abrikosov_2}
\end{equation}

\noindent The above conditions are generally satisfied either in materials with low carrier density ($\sim 10^{18}$ cm$^{-3}$) and small effective mass ($m^*$), such as bismuth (Bi)~\cite{Bi_1,Bi_2,Abrikosov_2000}, where $n_0 \sim 10^{17}$ cm$^{-3}$ and $m^* \sim 10^{-2}m_0$ ($m_0$ is the free-electron mass), or in materials with high carrier density, such as rare-earth diantimonides~\cite{Diantimonides,Abrikosov_2000}, when a large Fermi surface coexists with a small Fermi pocket associated with low effective mass~\cite{CrAs}. Our Hall effect analysis on MnP yields a carrier density of the order of $\sim 10^{22}$ cm$^{-3}$, consistent with the high carrier density reported for the related pnictide CrP ($\sim 10^{21}$ cm$^{-3}$)~\cite{CrP_MR}. Thus, MnP does not fall within the low-carrier-density regime.

Our first-principles calculations reveal the presence of a small Fermi pocket associated with the semi-Dirac-like crossing near the
high-symmetry point $Y$ in all three magnetic phases. The  linear branch of this semi-Dirac band is characterized by its band velocity
$v$, whereas the quadratic branch possesses a finite directional curvature effective mass $m^{*}$, as summarized in Table~\ref{tab:effective_mass_velocity}. In the following quantum-limit analysis, the effective mass entering the critical-field expression corresponds to the magnitude of curvature mass, $|m^{*}|$. The combination of a finite quadratic effective mass, a large characteristic velocity of the linear branch, and the small energy separation between the crossing and $E_F$ favors access to the extreme quantum-limit regime. Following Ref.~\cite{CrAs}, the critical magnetic field ($\mu_0H^c$) required to drive the small Fermi pocket into the extreme quantum limit can be estimated as

\begin{equation} \mu_0H^c = \frac{8\sqrt{2|m^{*}|}} {6\pi e\hbar v} [f(\gamma)]^{-3/2} \left[(E_F-E_c)+k_BT\right]^{3/2}, \end{equation}
where $|m^{*}|$ is the magnitude of the curvature effective mass extracted from the quadratic branch, $v$ is the characteristic band velocity obtained from the linear branch, and $E_c$ is the energy of the semi-Dirac-like crossing. The dimensionless function is defined as $f(\gamma)=(1+\gamma)^{2/3}-\gamma^{2/3}$, where $\gamma\in[0,1]$ is the phase factor.

\par
To obtain a quantitative estimate of $\mu_0H^c$ for MnP, we compare our results with those reported for CrAs~\cite{CrAs}, where $\mu_0H^c\approx4$ T for $E_F-E_c=3.7$ meV. Assuming the same value of $\gamma$ as in CrAs, the estimated $\mu_0H^c$ values for the three magnetic phases of MnP at $T=$ 0 and 4 K are listed in Table~\ref{tab:effective_mass_velocity}.

\par

It is evident from Table~\ref{tab:effective_mass_velocity} that the extreme quantum
limit is not reached in the SCR phase, as this phase exists only in the low field region. In the FAN phase, the estimated $\mu_0H^c$ is 3.47 T and 4.13 T at $T=$0 and 4 K, respectively. Our magnetization and transport measurements indicate that the FAN phase extends over approximately $0.85 \leq \mu_0H \leq 4.5$ T. Thus, although the FAN phase can reach the extreme quantum limit at sufficiently high fields, the observed MR may contain additional contributions from enhanced $s-d$ scattering because of its complex magnetic structure. This is also supported by the enhanced magnon contribution, $\alpha_{\mathrm{mag}}$, to $\rho_{xx}(T)$ in the FAN phase (Table~\ref{table:RT}). The absence of linear MR in the FAN phase may be attributed to the
coexistence of multiple scattering mechanisms, as also supported by the observed violation of Kohler's rule (see Appendix D). A particularly important consequence follows for the FM2 phase. Due to the very small Fermi pocket, the FM2 phase, which emerges above approximately 4.5 T, is expected to readily access the extreme quantum-limit regime, as the estimated critical fields are $\mu_0H^c\approx2.56$ T and 3.15 T at $T=0$ and 4 K, respectively. Moreover, the $s$-$d$ scattering is strongly suppressed in this phase as the magnetic moments become aligned along the applied field. Thus, once the FM2 phase is established, the applied field already exceeds the estimated quantum-limit threshold of the small Fermi pocket, while the reduced $s$-$d$ scattering favors the emergence of quantum linear MR within Abrikosov’s theory.

\section{Conclusion}
\par
In conclusion, we have presented a comprehensive experimental and theoretical investigation of the magneto-transport properties of helimagnetic MnP single crystals. Hall measurements reveal an AHE dominated by skew scattering in the high-temperature regime, while a finite THE emerges in the noncollinear FAN and low-temperature SCR phases. At low temperatures, MnP exhibits non-saturating positive MR that closely follow the magnetic phase evolution which vary linearly in the FM2 phase and reaches nearly 800\% at 4 K and 15 T. By combining systematic magnetotransport measurements with first-principles calculations, we identify a semi-Dirac-like band crossing near the $Y$ point whose energy progressively approaches the $E_F$ from the SCR to FAN and FM2 states. The resulting small Fermi pocket suggests that the FM2 phase can access the extreme quantum-limit regime at experimentally accessible fields, where the pocket is expected to enter the lowest-Landau-level regime, providing a natural microscopic framework for understanding the observed LPMR within Abrikosov’s quantum-MR theory. These results establish MnP as a model magnetic system in which magnetic-state evolution can directly tune low-energy electronic states and their associated quantum magnetotransport. 

\section{Acknowledgments}

P.C. gratefully acknowledges the DST-INSPIRE program (Grant No. DST/INSPIRE Fellowship/2019/IF190532) for research assistance. The access to the x-ray facilities in the Materials Characterization Laboratory at the ISIS Facility is gratefully acknowledged. The Kolkata Centre of UGC-DAE-CSR is acknowledged for transport studies. JS and MK acknowledge National Supercomputing Mission (NSM) for providing computing resources of ‘PARAM RUDRA’ at S.N. Bose National Centre for Basic Sciences, which is implemented by C-DAC and supported by the Ministry of Electronics and Information Technology (MeitY) and the Department of Science and Technology (DST), Government of India.

\appendix
\section{Experimental Techniques}
High quality single crystals of MnP, used in this study, were grown using the Sn-flux method~\cite{MnP_SC}. Initially, Mn and P powder and Sn shots were taken in the molar ratio of Mn:P:Sn = 1:1:10 in a alumina crucible and the vacuumed sealed in a quartz ampule. The quartz ampule was then slowly heated to 650$^{\circ}$ C, held there for 8 h, then heated up to 1100$^{\circ}$ C, held for 6 h, and then slowly cooled down to 600$^{\circ}$ C at a rate of 2$^{\circ}$ C/h. At this temperature, the ampule was removed and the liquid Sn flux was separated by centrifuge, yielding needle-like MnP single crystals.
\par
Crystal structure and phase purity of the sample were investigated through single-crystal XRD (SCXRD) at  295 K and 30 K. SCXRD data were collected using a Rigaku-Oxford diffraction Xtalab synergy single crystal diffractometer equipped with a HyPix hybrid pixel array detector using Mo-K$\alpha$ radiation. The SCXRD data reduction was done using CrysAlisPro software~\cite{CrysAlisPro2018}. The structural solution and refinement were performed using Jana2020 software~\cite{Jana2020}. Orientation of the crystal planes were identified using Rigaku x-ray diffractometer ($\lambda$ = 1.54 \AA). High-resolution transmission electron microscopy (HRTEM) and selected area electron diffraction (SAED) were performed using a JEOL TEM 2010. 
\par
Magnetic measurements were carried out using MPMS3 (SQUID Magnetometer) of Quantum Design. Hall and magneto-resistance (MR) measurements were performed on a cryogen-free high magnetic field system (Cryogenic Ltd. UK). Element specific X-ray absorption spectroscopy (XAS) and X-ray magnetic circular dichroism (XMCD) were performed in total electron yield (TEY) mode on I10 beamline at the Diamond Light Source, U.K.

\section{Computational Details}

First-principles electronic-structure calculations were carried out within the framework of density functional theory (DFT) using the Vienna \textit{ab initio} Simulation Package (VASP)~\cite{hafner2008ab}. Noncollinear spin-polarized density-functional-theory~\cite{hobbs2000fully} calculations were performed.
The exchange-correlation energy was treated within the generalized gradient approximation (GGA)~\cite{ziesche1998density}. The electronic wave functions were expanded in a plane-wave basis with a kinetic-energy cutoff of 600~eV, which was kept fixed for all magnetic configurations considered in this work. The Brillouin zone was sampled using a $9\times9\times3$ Monkhorst--Pack $k$-point mesh. The same numerical parameters were employed for the different magnetic states in order to enable a direct comparison of their electronic structures and to ensure that the observed changes near the Fermi energy originate from the magnetic configuration rather than from differences in the computational setup.

MnP crystallizes in the orthorhombic $Pnma$ structure and exhibits several magnetically distinct phases at low temperature and under an applied magnetic field. To describe the experimentally established magnetic structures within periodic first-principles calculations, enlarged magnetic supercells were constructed for the screw (SCR), fan (FAN), and field-polarized ferromagnetic (FM2) states. In the SCR phase, the Mn moments rotate within the $ab$ plane with a magnetic propagation vector directed along the crystallographic $c$ axis. A magnetic supercell containing nine crystallographic unit cells was therefore employed to represent the long-period noncollinear magnetic modulation. The corresponding magnetic arrangement is illustrated schematically in the inset of Fig.~\ref{fig:DFT} (a).

For the FAN phase, a magnetic supercell was constructed according to the previously reported fan-like spin arrangement~\cite{PhysRevLett.126.177205}. In this phase, the Mn moments remain predominantly within the $ab$ plane but oscillate about the field direction instead of undergoing the complete rotation characteristic of the SCR phase. The corresponding magnetic configuration is shown in the inset of Fig.~\ref{fig:DFT} (b). 

 A nine-unit-cell supercell was also employed for the FM2 calculation so that the electronic structures of the SCR and field-polarized states could be compared using equivalent supercell dimensions. In the FM2 state, the Mn moments were aligned according to the field-polarized ferromagnetic configuration as shown in the inset of Fig.~\ref{fig:DFT} (c).

The use of explicit magnetic supercells allows the modification of the electronic structure caused by the different noncollinear magnetic backgrounds to be incorporated directly into the DFT calculations. The electronic band structures were analyzed along the $S-Y-\Gamma$ path of the orthorhombic Brillouin zone. The notation of the high-symmetry points follows the convention of Bradley and Cracknell~\cite{BradleyCracknell}. The physical momentum coordinate used for the subsequent low-energy fitting was determined from the actual reciprocal-space displacement along the calculated path rather than from the discrete index of the sampled $k$ points. This allows the fitted band slopes and curvatures to be expressed in physical reciprocal-space units.

\section{Analysis of the Semi-Dirac Bands}

The quantity used to color the calculated electronic bands in Fig.~\ref{fig:DFT} of the main text is the local directional curvature effective mass, \begin{equation} m^{*}(k) = \frac{\hbar^{2}} {\displaystyle \frac{\partial^{2}E(k)} {\partial k^{2}}}, \end{equation} where $k$ denotes the physical reciprocal-space coordinate along the considered momentum direction. The sign of $m^{*}$ follows the local band curvature: $m^{*}>0$ corresponds to electron-like curvature, whereas $m^{*}<0$ corresponds to hole-like curvature. This signed local quantity is used for the color representation of the bands in Fig.~\ref{fig:DFT}. For the quantitative characterization of the semi-Dirac-like dispersion, only the low-energy states in the immediate vicinity of the $Y$ point and close to the Fermi energy were considered. The approximately linear branch was fitted using \begin{equation} E(q)=s q+c, \end{equation} where $q$ denotes the physical reciprocal-space displacement from the crossing and $s$ is the corresponding band slope. The characteristic band velocity was obtained as \begin{equation} v = \frac{1}{\hbar} \left| \frac{\partial E}{\partial q} \right| = \frac{|s|}{\hbar}. \end{equation} The branch showing quadratic dispersion was fitted using \begin{equation} E(q)=Aq^{2}+Bq+C, \end{equation} where $A$ is the quadratic coefficient. The corresponding directional curvature effective mass is then \begin{equation} m^{*} = \frac{\hbar^{2}} {\displaystyle \frac{\partial^{2}E}{\partial q^{2}}} = \frac{\hbar^{2}}{2A}. \end{equation} The quantity $|m^{*}|$ reported in Table~IV of the main text corresponds to the magnitude of the directional curvature effective mass extracted from the quadratic fit near the semi-Dirac-like crossing. Thus, $m^{*}(k)$ used in the color map of Fig.~\ref{fig:DFT} represents a local signed curvature quantity, whereas $|m^{*}|$ reported in Table~\ref{tab:effective_mass_velocity} represents the characteristic magnitude obtained from the low-energy quadratic fit. The band velocity $v$ reported in Table~\ref{tab:effective_mass_velocity} is independently obtained from the corresponding linear fit.

Using this procedure, the quadratic branches yield $|m^{*}|/m_e=0.68$, $0.70$, and $0.62$ for the SCR, FAN, and FM2 states, respectively. The corresponding characteristic velocities extracted from the linear branches are $8.9\times10^{4}$, $8.2\times10^{4}$, and $7.8\times10^{4}$~m~s$^{-1}$, respectively. These values are used in the quantum-limit analysis presented in the main text.

\begin{figure}[h!]
\centering
\includegraphics[width = 7 cm]{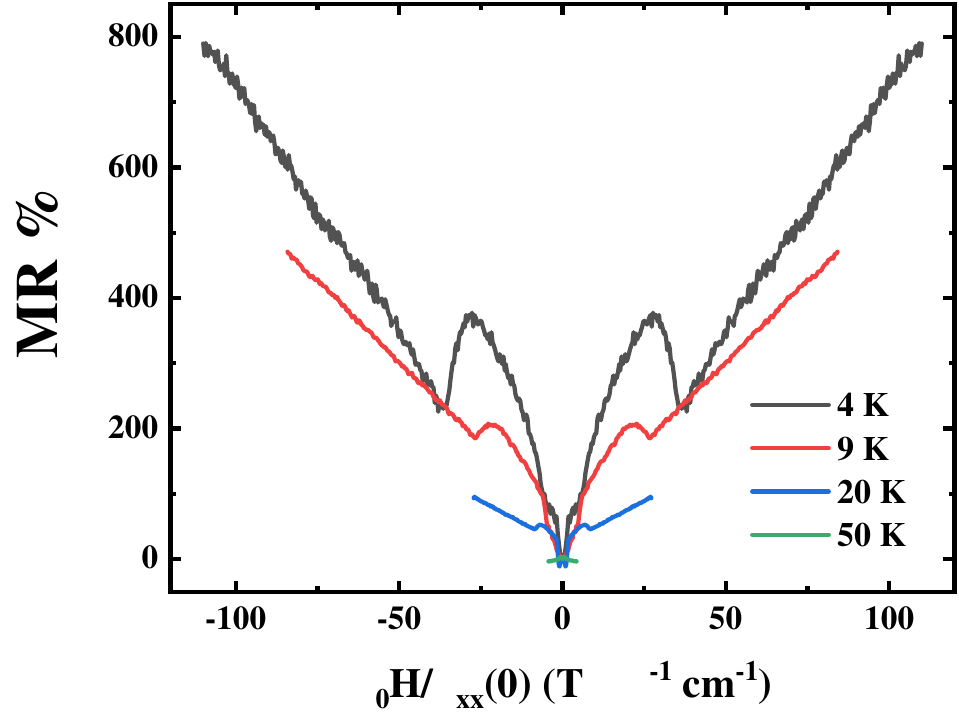}
\caption{Kohler’s scaling of MR at different temperatures ($T\leq50$ K).}
\label{fig:Kohler}
\end{figure}

\section{Violation of Kohler’s rule}

To investigate the underlying origin of the magnetoresistance, we plot MR\% as a function of $\mu_0H/\rho_{xx}(0)$ at different temperatures, as shown in Fig.~\ref{fig:Kohler}. The data do not collapse onto a single curve, indicating a violation of Kohler's rule, suggesting that more than one scattering mechanism contributes to the magnetotransport.

\bibliography{Reference}

@article{Jana2020,
url = {https://doi.org/10.1515/zkri-2023-0005},
title = {Jana2020– a new version of the crystallographic computing system Jana},
author = {Václav Petříček and Lukáš Palatinus and Jakub Plášil and Michal Dušek},
pages = {271--282},
volume = {238},
number = {7-8},
journal = {Zeitschrift für Kristallographie - Crystalline Materials},
doi = {doi:10.1515/zkri-2023-0005},
year = {2023},
lastchecked = {2025-03-20}
}

@article{T_star1,
doi = {10.1088/1742-6596/200/3/032079},
url = {https://dx.doi.org/10.1088/1742-6596/200/3/032079},
year = {2010},
month = {jan},
volume = {200},
number = {3},
pages = {032079},
author = {T Yamazaki and Y Tabata and T Waki and H Nakamura and M Matsuura and N Aso},
title = {Anomalous phase of MnP at very low field},
journal = {Journal of Physics: Conference Series}
}

@article{T_star2,
doi = {10.1088/0953-8984/12/27/307},
url = {https://dx.doi.org/10.1088/0953-8984/12/27/307},
year = {2000},
month = {jul},
volume = {12},
number = {27},
pages = {5889},
author = {C C Becerra},
title = {Evidence for a new magnetic phase in MnP at very low fields},
journal = {Journal of Physics: Condensed Matter}
}

@article{MnP_1966,
doi = {10.1088/0370-1328/88/2/308},
url = {https://dx.doi.org/10.1088/0370-1328/88/2/308},
year = {1966},
month = {jun},
publisher = {},
volume = {88},
number = {2},
pages = {333},
author = {J B Forsyth and S J Pickart and P J Brown},
title = {The structure of the metamagnetic phase of MnP},
journal = {Proceedings of the Physical Society}
}

@article{MnP_1980,
author = {Obara ,Hisashi and Endoh ,Yasuo and Ishikawa ,Yoshikazu and Komatsubara ,Takemi},
title = {Magnetic Phase Transition of MnP Under Magnetic Field},
journal = {Journal of the Physical Society of Japan},
volume = {49},
number = {3},
pages = {928-935},
year = {1980},
doi = {10.1143/JPSJ.49.928},

URL = { 
    
        https://doi.org/10.1143/JPSJ.49.928
    
    

}
}

@Article{FeP,
author={Campbell, D. J.
and Collini, J.
and S{\l}awi{\'{n}}ska, J.
and Autieri, C.
and Wang, L.
and Wang, K.
and Wilfong, B.
and Eo, Y. S.
and Neves, P.
and Graf, D.
and Rodriguez, E. E.
and Butch, N. P.
and Buongiorno Nardelli, M.
and Paglione, J.},
title={Topologically driven linear magnetoresistance in helimagnetic FeP},
journal={npj Quantum Materials},
year={2021},
month={Apr},
day={14},
volume={6},
number={1},
pages={38},
issn={2397-4648},
doi={10.1038/s41535-021-00337-2},
url={https://doi.org/10.1038/s41535-021-00337-2}
}

@Article{CrAs,
author={Niu, Q.
and Yu, W. C.
and Yip, K. Y.
and Lim, Z. L.
and Kotegawa, H.
and Matsuoka, E.
and Sugawara, H.
and Tou, H.
and Yanase, Y.
and Goh, Swee K.},
title={Quasilinear quantum magnetoresistance in pressure-induced nonsymmorphic superconductor chromium arsenide},
journal={Nature Communications},
year={2017},
month={Jun},
day={05},
volume={8},
number={1},
pages={15358},
issn={2041-1723},
doi={10.1038/ncomms15358},
url={https://doi.org/10.1038/ncomms15358}
}

@article{Fe3Sn2,
  title = {Anomalous Hall effect in a ferromagnetic ${\mathrm{Fe}}_{3}{\mathrm{Sn}}_{2}$ single crystal with a geometrically frustrated Fe bilayer kagome lattice},
  author = {Wang, Qi and Sun, Shanshan and Zhang, Xiao and Pang, Fei and Lei, Hechang},
  journal = {Phys. Rev. B},
  volume = {94},
  issue = {7},
  pages = {075135},
  numpages = {5},
  year = {2016},
  month = {Aug},
  publisher = {American Physical Society},
  doi = {10.1103/PhysRevB.94.075135},
  url = {https://link.aps.org/doi/10.1103/PhysRevB.94.075135}
}

@article{Cr2.76Te4,
  title = {Investigation of the anomalous and topological Hall effects in layered monoclinic ferromagnet ${\mathrm{Cr}}_{2.76}{\mathrm{Te}}_{4}$},
  author = {Purwar, Shubham and Low, Achintya and Bose, Anumita and Narayan, Awadhesh and Thirupathaiah, S.},
  journal = {Phys. Rev. Mater.},
  volume = {7},
  issue = {9},
  pages = {094204},
  numpages = {9},
  year = {2023},
  month = {Sep},
  publisher = {American Physical Society},
  doi = {10.1103/PhysRevMaterials.7.094204},
  url = {https://link.aps.org/doi/10.1103/PhysRevMaterials.7.094204}
}

@article{MnP_THE,
  title = {Emergence of topological Hall effect from fanlike spin structure as modified by Dzyaloshinsky-Moriya interaction in MnP},
  author = {Shiomi, Y. and Iguchi, S. and Tokura, Y.},
  journal = {Phys. Rev. B},
  volume = {86},
  issue = {18},
  pages = {180404},
  numpages = {4},
  year = {2012},
  month = {Nov},
  publisher = {American Physical Society},
  doi = {10.1103/PhysRevB.86.180404},
  url = {https://link.aps.org/doi/10.1103/PhysRevB.86.180404}
}

@article{MnP_JPSJ_Neutron,
author = {Yamazaki ,Teruo and Tabata ,Yoshikazu and Waki ,Takeshi and Sato ,Taku J. and Matsuura ,Masato and Ohoyama ,Kenji and Yokoyama ,Makoto and Nakamura ,Hiroyuki},
title = {Novel Magnetic Chiral Structures and Unusual Temperature Hysteresis in the Metallic Helimagnet MnP},
journal = {Journal of the Physical Society of Japan},
volume = {83},
number = {5},
pages = {054711},
year = {2014},
doi = {10.7566/JPSJ.83.054711},

URL = { 
    
        https://doi.org/10.7566/JPSJ.83.054711
    
    

}
}

@misc{CrysAlisPro2018,
  author       = {{Rigaku Oxford Diffraction}},
  year         = {2018},
  title        = {{CrysAlisPro Software System, Version 1.171.43.90}},
  howpublished = {Rigaku Corporation, Oxford, UK},
  url          = {https://www.rigakuxrayforum.com/forumdisplay.php?fid=57}
}

@article{THE_Thamizhavel,
  title = {Evidence of skyrmionic phase in a ${\mathrm{GdNi}}_{0.4}{\mathrm{Si}}_{1.6}$ single crystal},
  author = {Dwari, Gourav and Dan, Shovan and Maity, Bishal Baran and Kulkarni, Ruta and Thamizhavel, A.},
  journal = {Phys. Rev. B},
  volume = {110},
  issue = {13},
  pages = {134413},
  numpages = {7},
  year = {2024},
  month = {Oct},
  publisher = {American Physical Society},
  doi = {10.1103/PhysRevB.110.134413},
  url = {https://link.aps.org/doi/10.1103/PhysRevB.110.134413}
}

@Article{Chirality1,
author={Ueland, B. G.
and Miclea, C. F.
and Kato, Yasuyuki
and Ayala--Valenzuela, O.
and McDonald, R. D.
and Okazaki, R.
and Tobash, P. H.
and Torrez, M. A.
and Ronning, F.
and Movshovich, R.
and Fisk, Z.
and Bauer, E. D.
and Martin, Ivar
and Thompson, J. D.},
title={Controllable chirality-induced geometrical Hall effect in a frustrated highly correlated metal},
journal={Nature Communications},
year={2012},
month={Sep},
day={18},
volume={3},
number={1},
pages={1067},
issn={2041-1723},
doi={10.1038/ncomms2075},
url={https://doi.org/10.1038/ncomms2075}
}

@article{Chirality2,
  title = {Topological Hall Effect in Pyrochlore Lattice with Varying Density of Spin Chirality},
  author = {Ueda, K. and Iguchi, S. and Suzuki, T. and Ishiwata, S. and Taguchi, Y. and Tokura, Y.},
  journal = {Phys. Rev. Lett.},
  volume = {108},
  issue = {15},
  pages = {156601},
  numpages = {5},
  year = {2012},
  month = {Apr},
  publisher = {American Physical Society},
  doi = {10.1103/PhysRevLett.108.156601},
  url = {https://link.aps.org/doi/10.1103/PhysRevLett.108.156601}
}

@article{MnP_Helical,
    author = {Felcher, G. P.},
    title = {Magnetic Structure of MnP},
    journal = {Journal of Applied Physics},
    volume = {37},
    number = {3},
    pages = {1056-1058},
    year = {1966},
    month = {03},
    issn = {0021-8979},
    doi = {10.1063/1.1708333},
    url = {https://doi.org/10.1063/1.1708333},
    
}

@Article{CrAs_SC,
author={Wu, Wei
and Cheng, Jinguang
and Matsubayashi, Kazuyuki
and Kong, Panpan
and Lin, Fukun
and Jin, Changqing
and Wang, Nanlin
and Uwatoko, Yoshiya
and Luo, Jianlin},
title={Superconductivity in the vicinity of antiferromagnetic order in CrAs},
journal={Nature Communications},
year={2014},
month={Nov},
day={19},
volume={5},
number={1},
pages={5508},
issn={2041-1723},
doi={10.1038/ncomms6508},
url={https://doi.org/10.1038/ncomms6508}
}

@article{MnP_SC,
  title = {Pressure Induced Superconductivity on the border of Magnetic Order in MnP},
  author = {Cheng, J.-G. and Matsubayashi, K. and Wu, W. and Sun, J. P. and Lin, F. K. and Luo, J. L. and Uwatoko, Y.},
  journal = {Phys. Rev. Lett.},
  volume = {114},
  issue = {11},
  pages = {117001},
  numpages = {5},
  year = {2015},
  month = {Mar},
  publisher = {American Physical Society},
  doi = {10.1103/PhysRevLett.114.117001},
  url = {https://link.aps.org/doi/10.1103/PhysRevLett.114.117001}
}

@article{FeAs_NM1,
  title = {Non-nesting spin-density-wave antiferromagnetism in FeAs from first principles},
  author = {Parker, David and Mazin, I. I.},
  journal = {Phys. Rev. B},
  volume = {83},
  issue = {18},
  pages = {180403},
  numpages = {4},
  year = {2011},
  month = {May},
  publisher = {American Physical Society},
  doi = {10.1103/PhysRevB.83.180403},
  url = {https://link.aps.org/doi/10.1103/PhysRevB.83.180403}
}

@article{FeAs_NM2,
  title = {Quantum oscillations in the anomalous spin density wave state of FeAs},
  author = {Campbell, Daniel J. and Eckberg, Chris and Wang, Kefeng and Wang, Limin and Hodovanets, Halyna and Graf, Dave and Parker, David and Paglione, Johnpierre},
  journal = {Phys. Rev. B},
  volume = {96},
  issue = {7},
  pages = {075120},
  numpages = {7},
  year = {2017},
  month = {Aug},
  publisher = {American Physical Society},
  doi = {10.1103/PhysRevB.96.075120},
  url = {https://link.aps.org/doi/10.1103/PhysRevB.96.075120}
}

@article{CrAs_QC,
  title = {Evolution of Magnetic Double Helix and Quantum Criticality near a Dome of Superconductivity in CrAs},
  author = {Matsuda, M. and Lin, F. K. and Yu, R. and Cheng, J.-G. and Wu, W. and Sun, J. P. and Zhang, J. H. and Sun, P. J. and Matsubayashi, K. and Miyake, T. and Kato, T. and Yan, J.-Q. and Stone, M. B. and Si, Qimiao and Luo, J. L. and Uwatoko, Y.},
  journal = {Phys. Rev. X},
  volume = {8},
  issue = {3},
  pages = {031017},
  numpages = {12},
  year = {2018},
  month = {Jul},
  publisher = {American Physical Society},
  doi = {10.1103/PhysRevX.8.031017},
  url = {https://link.aps.org/doi/10.1103/PhysRevX.8.031017}
}

@article{CrP_MR,
  title = {Nonsaturating large magnetoresistance in the high carrier density nonsymmorphic metal CrP},
  author = {Niu, Q. and Yu, W. C. and Aulestia, E. I. Paredes and Hu, Y. J. and Lai, Kwing To and Kotegawa, H. and Matsuoka, E. and Sugawara, H. and Tou, H. and Sun, D. and Balakirev, F. F. and Yanase, Y. and Goh, Swee K.},
  journal = {Phys. Rev. B},
  volume = {99},
  issue = {12},
  pages = {125126},
  numpages = {6},
  year = {2019},
  month = {Mar},
  publisher = {American Physical Society},
  doi = {10.1103/PhysRevB.99.125126},
  url = {https://link.aps.org/doi/10.1103/PhysRevB.99.125126}
}

@article{Lifshitz_Point_MnP_PRL,
  title = {Lifshitz Point in MnP},
  author = {Becerra, C. C. and Shapira, Y. and Oliveira, N. F. and Chang, T. S.},
  journal = {Phys. Rev. Lett.},
  volume = {44},
  issue = {25},
  pages = {1692--1695},
  numpages = {0},
  year = {1980},
  month = {Jun},
  publisher = {American Physical Society},
  doi = {10.1103/PhysRevLett.44.1692},
  url = {https://link.aps.org/doi/10.1103/PhysRevLett.44.1692}
}

@article{MnP_Phase1,
    author = {Ishizaki, Ariyoshi and Komatsubara, Takemi and Hirahara, Eiji},
    title = {Magnetostriction in Manganese Phosphide Single Crystal},
    journal = {Progress of Theoretical Physics Supplement},
    volume = {46},
    pages = {256-279},
    year = {1970},
    month = {06},
    issn = {0375-9687},
    doi = {10.1143/PTPS.46.256},
    url = {https://doi.org/10.1143/PTPS.46.256},
    
}

@article{MnP_Phase2,
  title = {Phase transitions of MnP for a field parallel to the hard-magnetization direction: A possible new Lifshitz point},
  author = {Shapira, Y. and Oliveira, N. F. and Becerra, C. C. and Foner, S.},
  journal = {Phys. Rev. B},
  volume = {29},
  issue = {1},
  pages = {361--373},
  numpages = {0},
  year = {1984},
  month = {Jan},
  publisher = {American Physical Society},
  doi = {10.1103/PhysRevB.29.361},
  url = {https://link.aps.org/doi/10.1103/PhysRevB.29.361}
}

@article{Komatsubara,
author = {Komatsubara ,Takemi and Suzuki ,Takashi and Hirahara ,Eiji},
title = {Magnetization Process and Spin-Structure Diagram in Manganese Phosphide Single Crystal},
journal = {Journal of the Physical Society of Japan},
volume = {28},
number = {2},
pages = {317-320},
year = {1970},
doi = {10.1143/JPSJ.28.317},

URL = { 
    
        https://doi.org/10.1143/JPSJ.28.317
    
}
}

@article{MnP_Phase3,
  title = {Phase diagram, susceptibility, and magnetostriction of MnP: Evidence for a Lifshitz point},
  author = {Shapira, Y. and Becerra, C. C. and Oliveira, N. F. and Chang, T. S.},
  journal = {Phys. Rev. B},
  volume = {24},
  issue = {5},
  pages = {2780--2806},
  numpages = {0},
  year = {1981},
  month = {Sep},
  publisher = {American Physical Society},
  doi = {10.1103/PhysRevB.24.2780},
  url = {https://link.aps.org/doi/10.1103/PhysRevB.24.2780}
}

@article{XMCD_rule1,
  title = {X-ray circular dichroism as a probe of orbital magnetization},
  author = {Thole, B. T. and Carra, P. and Sette, F. and van der Laan, G.},
  journal = {Phys. Rev. Lett.},
  volume = {68},
  issue = {12},
  pages = {1943--1946},
  numpages = {0},
  year = {1992},
  month = {Mar},
  publisher = {American Physical Society},
  doi = {10.1103/PhysRevLett.68.1943},
  url = {https://link.aps.org/doi/10.1103/PhysRevLett.68.1943}
}

@article{XMCD_rule2,
  title = {Experimental Confirmation of the X-Ray Magnetic Circular Dichroism Sum Rules for Iron and Cobalt},
  author = {Chen, C. T. and Idzerda, Y. U. and Lin, H.-J. and Smith, N. V. and Meigs, G. and Chaban, E. and Ho, G. H. and Pellegrin, E. and Sette, F.},
  journal = {Phys. Rev. Lett.},
  volume = {75},
  issue = {1},
  pages = {152--155},
  numpages = {0},
  year = {1995},
  month = {Jul},
  publisher = {American Physical Society},
  doi = {10.1103/PhysRevLett.75.152},
  url = {https://link.aps.org/doi/10.1103/PhysRevLett.75.152}
}

@article{XMCD_Mn2VAl,
  title = {Electronic structure and magnetic properties of the half-metallic ferrimagnet ${\mathrm{Mn}}_{2}\mathrm{VAl}$ probed by soft x-ray spectroscopies},
  author = {Nagai, K. and Fujiwara, H. and Aratani, H. and Fujioka, S. and Yomosa, H. and Nakatani, Y. and Kiss, T. and Sekiyama, A. and Kuroda, F. and Fujii, H. and Oguchi, T. and Tanaka, A. and Miyawaki, J. and Harada, Y. and Takeda, Y. and Saitoh, Y. and Suga, S. and Umetsu, R. Y.},
  journal = {Phys. Rev. B},
  volume = {97},
  issue = {3},
  pages = {035143},
  numpages = {8},
  year = {2018},
  month = {Jan},
  publisher = {American Physical Society},
  doi = {10.1103/PhysRevB.97.035143},
  url = {https://link.aps.org/doi/10.1103/PhysRevB.97.035143}
}

@article{XMCD_Mn_Film,
  title = {Electron-correlation-induced magnetic order of ultrathin Mn films},
  author = {D\"urr, H. A. and van der Laan, G. and Spanke, D. and Hillebrecht, F. U. and Brookes, N. B.},
  journal = {Phys. Rev. B},
  volume = {56},
  issue = {13},
  pages = {8156--8162},
  numpages = {0},
  year = {1997},
  month = {Oct},
  publisher = {American Physical Society},
  doi = {10.1103/PhysRevB.56.8156},
  url = {https://link.aps.org/doi/10.1103/PhysRevB.56.8156}
}

@Article{XMCD_Mn3Sn,
author={Kimata, Motoi
and Sasabe, Norimasa
and Kurita, Kensuke
and Yamasaki, Yuichi
and Tabata, Chihiro
and Yokoyama, Yuichi
and Kotani, Yoshinori
and Ikhlas, Muhammad
and Tomita, Takahiro
and Amemiya, Kenta
and Nojiri, Hiroyuki
and Nakatsuji, Satoru
and Koretsune, Takashi
and Nakao, Hironori
and Arima, Taka-hisa
and Nakamura, Tetsuya},
title={X-ray study of ferroic octupole order producing anomalous Hall effect},
journal={Nature Communications},
year={2021},
month={Sep},
day={22},
volume={12},
number={1},
pages={5582},
issn={2041-1723},
doi={10.1038/s41467-021-25834-7},
url={https://doi.org/10.1038/s41467-021-25834-7}
}

@ARTICLE{MR_appl1,
  author={Wang, Shan X. and Li, Guanxiong},
  journal={IEEE Transactions on Magnetics}, 
  title={Advances in Giant Magnetoresistance Biosensors With Magnetic Nanoparticle Tags: Review and Outlook}, 
  year={2008},
  volume={44},
  number={7},
  pages={1687-1702},
  doi={10.1109/TMAG.2008.920962}}

@article{MR_appl2,
  title = {GMR applications},
journal = {Journal of Magnetism and Magnetic Materials},
volume = {192},
number = {2},
pages = {334-342},
year = {1999},
issn = {0304-8853},
doi = {https://doi.org/10.1016/S0304-8853(98)00376-X},
url = {https://www.sciencedirect.com/science/article/pii/S030488539800376X},
author = {J.M. Daughton}
}

@Article{WTe2,
author={Ali, Mazhar N.
and Xiong, Jun
and Flynn, Steven
and Tao, Jing
and Gibson, Quinn D.
and Schoop, Leslie M.
and Liang, Tian
and Haldolaarachchige, Neel
and Hirschberger, Max
and Ong, N. P.
and Cava, R. J.},
title={Large, non-saturating magnetoresistance in WTe2},
journal={Nature},
year={2014},
month={Oct},
day={01},
volume={514},
number={7521},
pages={205-208},
issn={1476-4687},
doi={10.1038/nature13763},
url={https://doi.org/10.1038/nature13763}
}

@Article{NbP,
author={Shekhar, Chandra
and Nayak, Ajaya K.
and Sun, Yan
and Schmidt, Marcus
and Nicklas, Michael
and Leermakers, Inge
and Zeitler, Uli
and Skourski, Yurii
and Wosnitza, Jochen
and Liu, Zhongkai
and Chen, Yulin
and Schnelle, Walter
and Borrmann, Horst
and Grin, Yuri
and Felser, Claudia
and Yan, Binghai},
title={Extremely large magnetoresistance and ultrahigh mobility in the topological Weyl semimetal candidate NbP},
journal={Nature Physics},
year={2015},
month={Aug},
day={01},
volume={11},
number={8},
pages={645-649},
issn={1745-2481},
doi={10.1038/nphys3372},
url={https://doi.org/10.1038/nphys3372}
}

@Article{MnBi,
author={He, Yangkun
and Gayles, Jacob
and Yao, Mengyu
and Helm, Toni
and Reimann, Tommy
and Strocov, Vladimir N.
and Schnelle, Walter
and Nicklas, Michael
and Sun, Yan
and Fecher, Gerhard H.
and Felser, Claudia},
title={Large linear non-saturating magnetoresistance and high mobility in ferromagnetic MnBi},
journal={Nature Communications},
year={2021},
month={Jul},
day={28},
volume={12},
number={1},
pages={4576},
issn={2041-1723},
doi={10.1038/s41467-021-24692-7},
url={https://doi.org/10.1038/s41467-021-24692-7}
}

@article{Cd3As2,
  title = {Large linear magnetoresistance in Dirac semimetal ${\mathrm{Cd}}_{3}{\mathrm{As}}_{2}$ with Fermi surfaces close to the Dirac points},
  author = {Feng, Junya and Pang, Yuan and Wu, Desheng and Wang, Zhijun and Weng, Hongming and Li, Jianqi and Dai, Xi and Fang, Zhong and Shi, Youguo and Lu, Li},
  journal = {Phys. Rev. B},
  volume = {92},
  issue = {8},
  pages = {081306(R)},
  numpages = {5},
  year = {2015},
  month = {Aug},
  publisher = {American Physical Society},
  doi = {10.1103/PhysRevB.92.081306},
  url = {https://link.aps.org/doi/10.1103/PhysRevB.92.081306}
}

@article{ZrSiS,
author = {Ratnadwip Singha  and Arnab Kumar Pariari  and Biswarup Satpati  and Prabhat Mandal },
title = {Large nonsaturating magnetoresistance and signature of nondegenerate Dirac nodes in ZrSiS},
journal = {Proceedings of the National Academy of Sciences},
volume = {114},
number = {10},
pages = {2468-2473},
year = {2017},
doi = {10.1073/pnas.1618004114},
URL = {https://www.pnas.org/doi/abs/10.1073/pnas.1618004114}}

@article{CoS2_PNAS,
author = {Shen Zhang  and Yibo Wang  and Qingqi Zeng  and Jianlei Shen  and Xinqi Zheng  and Jinying Yang  and Zhaosheng Wang  and Chuanying Xi  and Binbin Wang  and Min Zhou  and Rongjin Huang  and Hongxiang Wei  and Yuan Yao  and Shouguo Wang  and Stuart S. P. Parkin  and Claudia Felser  and Enke Liu  and Baogen Shen },
title = {Scaling of Berry-curvature monopole dominated large linear positive magnetoresistance},
journal = {Proceedings of the National Academy of Sciences},
volume = {119},
number = {45},
pages = {e2208505119},
year = {2022},
doi = {10.1073/pnas.2208505119},
URL = {https://www.pnas.org/doi/abs/10.1073/pnas.2208505119}}

@article{PtBi2,
  title = {Extremely Large Magnetoresistance in a Topological Semimetal Candidate Pyrite ${\mathrm{PtBi}}_{2}$},
  author = {Gao, Wenshuai and Hao, Ningning and Zheng, Fa-Wei and Ning, Wei and Wu, Min and Zhu, Xiangde and Zheng, Guolin and Zhang, Jinglei and Lu, Jianwei and Zhang, Hongwei and Xi, Chuanying and Yang, Jiyong and Du, Haifeng and Zhang, Ping and Zhang, Yuheng and Tian, Mingliang},
  journal = {Phys. Rev. Lett.},
  volume = {118},
  issue = {25},
  pages = {256601},
  numpages = {5},
  year = {2017},
  month = {Jun},
  publisher = {American Physical Society},
  doi = {10.1103/PhysRevLett.118.256601},
  url = {https://link.aps.org/doi/10.1103/PhysRevLett.118.256601}
}

@Article{Ag2Se,
author={Xu, R.
and Husmann, A.
and Rosenbaum, T. F.
and Saboungi, M.-L.
and Enderby, J. E.
and Littlewood, P. B.},
title={Large magnetoresistance in non-magnetic silver chalcogenides},
journal={Nature},
year={1997},
month={Nov},
day={01},
volume={390},
number={6655},
pages={57-60},
issn={1476-4687},
doi={10.1038/36306},
url={https://doi.org/10.1038/36306}
}

@article{Cd3As2_PRL,
  title = {Linear Magnetoresistance Caused by Mobility Fluctuations in $n$-Doped ${\mathrm{Cd}}_{3}{\mathrm{As}}_{2}$},
  author = {Narayanan, A. and Watson, M. D. and Blake, S. F. and Bruyant, N. and Drigo, L. and Chen, Y. L. and Prabhakaran, D. and Yan, B. and Felser, C. and Kong, T. and Canfield, P. C. and Coldea, A. I.},
  journal = {Phys. Rev. Lett.},
  volume = {114},
  issue = {11},
  pages = {117201},
  numpages = {5},
  year = {2015},
  month = {Mar},
  publisher = {American Physical Society},
  doi = {10.1103/PhysRevLett.114.117201},
  url = {https://link.aps.org/doi/10.1103/PhysRevLett.114.117201}
}

@article{TlBiSSe_LMR,
  title = {Large linear magnetoresistance in the Dirac semimetal TlBiSSe},
  author = {Novak, Mario and Sasaki, Satoshi and Segawa, Kouji and Ando, Yoichi},
  journal = {Phys. Rev. B},
  volume = {91},
  issue = {4},
  pages = {041203(R)},
  numpages = {5},
  year = {2015},
  month = {Jan},
  publisher = {American Physical Society},
  doi = {10.1103/PhysRevB.91.041203},
  url = {https://link.aps.org/doi/10.1103/PhysRevB.91.041203}
}

@article{CrTe_PRM,
  title = {Suppression of intrinsic Hall effect through competing Berry curvature in ${\mathrm{Cr}}_{1+\ensuremath{\delta}}{\mathrm{Te}}_{2}$},
  author = {Chowdhury, Prasanta and Sau, Jyotirmay and Numan, Mohamad and Sannigrahi, Jhuma and Gutmann, Matthias and Giri, Saurav and Kumar, Manoranjan and Majumdar, Subham},
  journal = {Phys. Rev. Mater.},
  volume = {9},
  issue = {2},
  pages = {024407},
  numpages = {12},
  year = {2025},
  month = {Feb},
  publisher = {American Physical Society},
  doi = {10.1103/PhysRevMaterials.9.024407},
  url = {https://link.aps.org/doi/10.1103/PhysRevMaterials.9.024407}
}

@article{SmMn2Ge2,
  title = {Linear magnetoresistance, anomalous Hall effect, and de Haas--van Alphen oscillations in antiferromagnetic ${\mathrm{SmAg}}_{2}{\mathrm{Ge}}_{2}$ single crystals},
  author = {Bala, Kanchan and Verma, Rahul and Dan, Shovan and Nandi, Suman and Kulkarni, Ruta and Singh, Bahadur and Thamizhavel, A.},
  journal = {Phys. Rev. B},
  volume = {111},
  issue = {24},
  pages = {245123},
  numpages = {12},
  year = {2025},
  month = {Jun},
  publisher = {American Physical Society},
  doi = {10.1103/PhysRevB.111.245123},
  url = {https://link.aps.org/doi/10.1103/PhysRevB.111.245123}
}

@Article{PL_model,
author={Parish, M. M.
and Littlewood, P. B.},
title={Non-saturating magnetoresistance in heavily disordered semiconductors},
journal={Nature},
year={2003},
month={Nov},
day={01},
volume={426},
number={6963},
pages={162-165},
issn={1476-4687},
doi={10.1038/nature02073},
url={https://doi.org/10.1038/nature02073}
}

@article{Abrikosov_2000,
doi = {10.1209/epl/i2000-00220-2},
url = {https://doi.org/10.1209/epl/i2000-00220-2},
year = {2000},
month = {mar},
publisher = {},
volume = {49},
number = {6},
pages = {789},
author = {A. A. Abrikosov},
title = {Quantum linear magnetoresistance},
journal = {Europhysics Letters}
}

@article{Abrikosov_1998,
  title = {Quantum magnetoresistance},
  author = {Abrikosov, A. A.},
  journal = {Phys. Rev. B},
  volume = {58},
  issue = {5},
  pages = {2788--2794},
  numpages = {0},
  year = {1998},
  month = {Aug},
  publisher = {American Physical Society},
  doi = {10.1103/PhysRevB.58.2788},
  url = {https://link.aps.org/doi/10.1103/PhysRevB.58.2788}
}

@article{Bi_1,
    author = {Kapitza, P.},
    title = {The study of the specific resistance of bismuth crystals and its change in strong magnetic fields and some allied problems},
    journal = {Proceedings of the Royal Society of London. Series A, Containing Papers of a Mathematical and Physical Character},
    volume = {119},
    number = {782},
    pages = {358-443},
    year = {1928},
    month = {06},
    issn = {0950-1207},
    doi = {10.1098/rspa.1928.0103},
    url = {https://doi.org/10.1098/rspa.1928.0103},
}

@article{Bi_2,
author = {F. Y. Yang  and Kai Liu  and Kimin Hong  and D. H. Reich  and P. C. Searson  and C. L. Chien },
title = {Large Magnetoresistance of Electrodeposited Single-Crystal Bismuth Thin Films},
journal = {Science},
volume = {284},
number = {5418},
pages = {1335-1337},
year = {1999},
doi = {10.1126/science.284.5418.1335},
URL = {https://www.science.org/doi/abs/10.1126/science.284.5418.1335}}

@article{Diantimonides,
  title = {Anisotropic magnetic properties of light rare-earth diantimonides},
  author = {Bud'ko, S. L. and Canfield, P. C. and Mielke, C. H. and Lacerda, A. H.},
  journal = {Phys. Rev. B},
  volume = {57},
  issue = {21},
  pages = {13624--13638},
  numpages = {0},
  year = {1998},
  month = {Jun},
  publisher = {American Physical Society},
  doi = {10.1103/PhysRevB.57.13624},
  url = {https://link.aps.org/doi/10.1103/PhysRevB.57.13624}
}

@article{PhysRevLett.126.177205,
  title = {Chirality Memory Stored in Magnetic Domain Walls in the Ferromagnetic State of MnP},
  author = {Jiang, N. and Nii, Y. and Arisawa, H. and Saitoh, E. and Ohe, J. and Onose, Y.},
  journal = {Phys. Rev. Lett.},
  volume = {126},
  issue = {17},
  pages = {177205},
  numpages = {5},
  year = {2021},
  month = {Apr},
  publisher = {American Physical Society},
  doi = {10.1103/PhysRevLett.126.177205},
  url = {https://link.aps.org/doi/10.1103/PhysRevLett.126.177205}
}

@book{BradleyCracknell, author = {C. J. Bradley and A. P. Cracknell}, title = {The Mathematical Theory of Symmetry in Solids: Representation Theory for Point Groups and Space Groups}, publisher = {Oxford University Press}, address = {Oxford}, year = {2009} }

@article{hafner2008ab,
  title={Ab-initio simulations of materials using VASP: Density-functional theory and beyond},
  author={Hafner, J{\"u}rgen},
  journal={Journal of computational chemistry},
  volume={29},
  number={13},
  pages={2044--2078},
  year={2008},
  publisher={Wiley Online Library}
}

@article{ziesche1998density,
  title={Density functionals from LDA to GGA},
  author={Ziesche, Paul and Kurth, Stefan and Perdew, John P},
  journal={Computational materials science},
  volume={11},
  number={2},
  pages={122--127},
  year={1998},
  publisher={Elsevier}
}

@article{hobbs2000fully,
  title={Fully unconstrained noncollinear magnetism within the projector augmented-wave method},
  author={Hobbs, D and Kresse, G and Hafner, J},
  journal={Physical Review B},
  volume={62},
  number={17},
  pages={11556},
  year={2000},
  publisher={APS}
}

@article{suzuki1995first,
  title={First-principles calculations of effective-mass parameters of AlN and GaN},
  author={Suzuki, Masakatsu and Uenoyama, Takeshi and Yanase, Akira},
  journal={Physical Review B},
  volume={52},
  number={11},
  pages={8132},
  year={1995},
  publisher={APS}
}

@Article{Mn5Si3_THE,
author={S{\"u}rgers, Christoph
and Fischer, Gerda
and Winkel, Patrick
and L{\"o}hneysen, Hilbert v.},
title={Large topological Hall effect in the non-collinear phase of an antiferromagnet},
journal={Nature Communications},
year={2014},
month={Mar},
day={05},
volume={5},
number={1},
pages={3400},
issn={2041-1723},
doi={10.1038/ncomms4400},
url={https://doi.org/10.1038/ncomms4400}
}

@article{SmMn2Ge2_THE,
  title = {Multiple unconventional Hall effects induced by noncoplanar spin textures in $\mathrm{SmM}{\mathrm{n}}_{2}\mathrm{G}{\mathrm{e}}_{2}$},
  author = {Huang, Dan and Li, Hang and Ding, Bei and Song, Linxuan and Li, Xue and Xi, Xuekui and Lau, Yong-Chang and Gao, Jianrong and Wang, Wenhong},
  journal = {Phys. Rev. B},
  volume = {109},
  issue = {14},
  pages = {144406},
  numpages = {12},
  year = {2024},
  month = {Apr},
  publisher = {American Physical Society},
  doi = {10.1103/PhysRevB.109.144406},
  url = {https://link.aps.org/doi/10.1103/PhysRevB.109.144406}
}

@article{Nagaosa1,
  title = {Intrinsic Versus Extrinsic Anomalous Hall Effect in Ferromagnets},
  author = {Onoda, Shigeki and Sugimoto, Naoyuki and Nagaosa, Naoto},
  journal = {Phys. Rev. Lett.},
  volume = {97},
  issue = {12},
  pages = {126602},
  numpages = {4},
  year = {2006},
  month = {Sep},
  publisher = {American Physical Society},
  doi = {10.1103/PhysRevLett.97.126602},
  url = {https://link.aps.org/doi/10.1103/PhysRevLett.97.126602}
}

@article{Nagaosa2,
  title = {Anomalous Hall effect},
  author = {Nagaosa, Naoto and Sinova, Jairo and Onoda, Shigeki and MacDonald, A. H. and Ong, N. P.},
  journal = {Rev. Mod. Phys.},
  volume = {82},
  issue = {2},
  pages = {1539--1592},
  numpages = {0},
  year = {2010},
  month = {May},
  publisher = {American Physical Society},
  doi = {10.1103/RevModPhys.82.1539},
  url = {https://link.aps.org/doi/10.1103/RevModPhys.82.1539}
}

\end{document}